\documentclass[12pt,a4j]{article}
\usepackage[dvipdfmx]{color,graphicx}
\usepackage{amsmath,enumerate, amsfonts, ulem, url}
\usepackage{natbib}
\newcommand{\bm}[1]{\mbox{\boldmath{$#1 $}}}

\begin{document}
	\baselineskip 18pt
	\begin{center}
		{\LARGE\textbf{Quadratic regression for functional response models}}
	\end{center}
	\begin{center}
		{\large Hidetoshi Matsui}
	\end{center}
	
	\begin{center}
		\begin{minipage}{14cm}
			{
				\begin{center}
					{\it {\footnotesize 
							Faculty of Data Science, Shiga University \\
							1-1-1, Banba, Hikone, Shiga 522-8522, Japan. \\
					}}
					
					\vspace{2mm}
					
					{\small hmatsui@biwako.shiga-u.ac.jp}
				\end{center}
				\vspace{1mm} 
				
				{\small {\bf Abstract:} 
					We consider the problem of constructing a regression model with a functional predictor and a functional response.  
					We extend the functional linear model to the quadratic model, where the quadratic term also takes the interaction between the argument of the functional data  into consideration.
					We assume that the predictor and the coefficient functions are expressed by basis expansions, and then parameters included in the model are estimated by the penalized likelihood method assuming that the error function follows a Gaussian process.  
					Monte Carlo simulations are conducted to illustrate the efficacy of the proposed method.  
					Finally, we apply the proposed method to the analysis of meteorological data and explore the results.
				}
				
				\vspace{3mm}
				
				{\small \noindent {\bf Key Words and Phrases:} Functional data analysis; Gaussian process; Interaction }
			}
		\end{minipage}
	\end{center}
\section{Introduction}
Functional data analysis (FDA) has received considerable attentions in various fields of application such as bioscience and meteorology. 
A number of methodological, theoretical, and empirical developments have occured over time which have improved and extended this technique \citep{RaSi2005,HoKo2012,KoRe2017}.  
The basic idea behind FDA is to express each individual in repeatedly measured data as a smooth function and then draw information from the collection of functional data.  
We consider the problem of constructing a regression model where both a predictor and the response are given as functional data.  

Functional regression analysis has been widely studied in the literature.  
When the predictor is a function but the response is a scalar, \cite{Ja2002} considered such models in the framework of generalized linear models.  
In addition, many other extensions such as adaptive models and neural networks have been reported \citep{MuSt2005,RoDeCo_etal2005}.  
Although flexible models capture the complex structures representing the relationships among variables, the results can be difficult to mechanistically interpret.  
\cite{YaMu2010} proposed a functional regression model with a quadratic term of the functional predictor.  
Suppose we have a functional predictor $X(t)$ and a scalar response $Y$; then the functional quadratic model is given by 
\begin{align*}
Y= \alpha + \int X(t)\beta(t)dt + \iint X(s)X(t)\gamma(s,t)dsdt + \varepsilon,
\end{align*}
where $\alpha$ is an intercept; $\beta(t)$ and $\gamma(s,t)$ are coefficient surfaces for linear and quadratic terms, respectively; and $\varepsilon$ is an additive error term.
Therefore, this model accommodate an interaction between a functional predictor $X$ at two arbitrary different time points.    
\cite{FuScGr2015} and \cite{UsStMa2016} proposed functional regression models that consider interactions between multiple functional predictors.  
\cite{WeTaLi2014} constructed a functional logistic regression model that considers an interaction between genetic variants and repeatedly measured environmental exposure to classify the disease status of patients.  
Their results suggested that disease classification was sensitive to the modeled variables and their interactions.  

Functional regression models with a functional predictor and a functional response are also considered in \cite{RaDa1991}, \cite{YaMuWa2005b}, and \cite{ScStGr2015}.  
This type of model is a useful tool for simulating the relationship between predictor and response functions at arbitrary times.  
However, hitherto, neither quadratic nor interaction terms have been considered with this type of model.  
In this paper, we introduce a quadratic term into the functional regression model with a functional predictor and a functional response.  
The predictor and coefficient functions are expressed by basis expansions; we show that, from these assumptions the problem of estimating coefficient functions becomes that of estimating parameter vectors.  
Furthermore, we consider estimating these parameters by the penalized likelihood method, where we assume the error function follows a Gaussian process, based on \cite{ShCh2011}.
Values of tuning parameters included in the estimation process are determined by model selection criteria rooted in information theory and a Bayesian approach \citep{KoKi2008}.
We also illustrate the efficacy of our method through Monte Carlo simulations and empirical analysis of meteorological data.  

The remainder of the paper is organized as follows.  
Section 2 specifies a functional quadratic model for a functional predictor and a functional response. 
In Section 3, we introduce the method for model estimation and evaluation in a maximum likelihood framework.  
We report the results of simulation studies in Section 4 before applying the proposed method to the analysis of empirical data in Section 5.  
Finally Section 6 offers conclusions.  

\section{Functional quadratic model}
Suppose we have $n$ sets of a functional predictor and a functional response $\{(x_i(s), y_i(t)); i=1,\ldots, n, s\in\mathcal S\subset\mathbb R, t\in\mathcal T \in \mathbb R \}$. 
We model the relationship between the predictor $x_i$ and the response $y_i$ as the following functional quadratic model with interactions as follows: 
\begin{align}
	y_i(t)= \alpha(t) +
	\int_{\mathcal S}x_{i}(s)\beta(s,t)ds + 
	{\iint_{\mathcal S\times \mathcal S}x_{i}(r)x_i(s)\gamma(r,s,t)drds} + 
	\varepsilon_i(t),
	\label{eq:model}
\end{align}
where $\alpha(t)$ is a baseline function, $\beta(s,t)$ is the coefficient surface for the linear term, $\gamma(r,s,t)$ is the coefficient hypersurface for the quadratic term, and $\varepsilon_i(t)$ is an error term.
Each coefficient function explains the weight of the predictor on the response at different time points.  
In particular, the quadratic term considers the interaction of $x_i$ at two  different time points $r$ and $s$ and $\gamma(r,s,t)$ represent weights of these interactions on the response function.  

To estimate coefficient functions in functional regression models, several methods have been proposed.  
Here we apply the basis expansion method, that is, the predictors $x_i(s)$ are expressed by linear combinations of basis functions:   
\begin{align}
x_i(s) = \sum_{k=1}^{M_x} w_{ik}\phi_k(s) = \bm w_i^T\bm{\phi}(s),
\label{eq:basis_x}
\end{align}
where $\bm{\phi}(s) = (\phi_{1}(s), \ldots, \phi_{M_x}(s))^T$ is a vector of $M_x$ basis functions and $\bm w_{i} = (w_{i1},\ldots, w_{iM_x})^T$ is a vector of weights that are obtained using smoothing techniques \citep[see, e.g.,][]{GrSi1994, ArKoKa_etal2009a}.  
In addition, we assume that the baseline and coefficient functions are also expressed by basis expansions as follows:
\begin{align}
&\alpha(t) = \sum_{l=1}^{M_y} a_l\psi_l(t) = \bm a^T\bm{\psi}(t),\nonumber\\
&\beta(s,t) = \sum_{k=1}^{M_x}\sum_{l=1}^{M_y} b_{kl}\phi_k(s)\psi_l(t) 
= \bm\phi(s)^T B\bm{\psi}(t),\label{eq:basis_b}\\
&\gamma(r,s,t) = \sum_{h=1}^{M_x}\sum_{k=1}^{M_x}\sum_{l=1}^{M_y} \gamma_{hkl}\phi_h(r)\phi_k(s)\psi_l(t) 
= \left\{\bm{\phi}(s)\otimes \bm{\phi}(r) \right\}^T {\Gamma}^T_{(3)} \bm{\psi}(t),\nonumber
\end{align}
where $\bm{\psi}(t) = (\psi_{1}(t), \ldots, \psi_{M_y}(t))^T$ is a vector of $M_y$ basis functions, $\bm{\alpha} = (\alpha_1,\ldots, \alpha_{M_y})^T$, $B=(b_{kl})_{kl}$ and $\Gamma_{(3)}$ is an $M_y \times M_x^2$ matrix obtained by matricizing a 3-dimensional $M_x \times M_x \times M_y$ tensor $\underline{\Gamma}=(\gamma_{hkl})_{hkl}$ with respect to the 3rd array.  
The strict definition of $\Gamma_{(3)}$ is used following \cite{DeDeVa2000}.  
In general, the functional regression model with basis expansions reduces the number of model parameters more than the traditional regression model, because the number of basis functions is smaller than the number of time points.  
This leads to model dimension reduction and provides more stable estimates.  
Using the above assumptions, the functional quadratic model (\ref{eq:model}) is written as
\begin{align*}
y_i(t) &= \bm a^T\bm{\psi}(t) + \bm w_i^T\Phi B\bm{\psi}(t) + (\bm w_i \otimes \bm w_i)^T (\Phi\otimes \Phi)\Gamma_{(3)}^T \bm{\psi}(t) + \varepsilon_i(t)\\
&= \bm z_i^T\Theta^T \bm\psi(t) + \varepsilon_i(t),
\end{align*}
where $\Phi = \int \bm{\phi}(s)\bm{\phi}(s)^Tds$, $\bm z_i = (1,\bm w_i^T\Phi, (\bm w_i \otimes \bm w_i)^T (\Phi\otimes \Phi))^T$ and $\Theta = (\bm{\alpha}~B^T~\Gamma_{(3)})^T$ is a matrix of parameters.  
Thus, the problem of estimating the baseline and coefficient functions becomes one of estimating the parameter matrix $\Theta$.   

More generally, when we have a $p$-th order interaction term with respect to $x_i$ in the functional regression model, using the same assumptions described above, this term is expressed by
\begin{align*}
	&\int\cdots\int_{\mathcal S^p}x_{i}(s_1)\cdots x_i(s_p)\gamma(s_1,\ldots, s_p, t)ds_1\cdots ds_p\\
	&= \int\cdots\int_{\mathcal S^p} 
	(\bm w_i \otimes \cdots \otimes \bm w_i)^T
	(\bm{\phi}(s_p)\otimes \cdots \otimes \bm{\phi}(s_1))\\
	& \hspace{2cm}
	\times (\bm{\phi}(s_p)\otimes \cdots \otimes \bm{\phi}(s_1))^T
	\Gamma_{(p+1)}^T\bm{\psi}(t)
	ds_1\cdots ds_p\\
	&= (\bm w_i \otimes \cdots \otimes \bm w_i)^T (\Phi\otimes \cdots \otimes \Phi)\Gamma_{(p+1)}^T \bm{\psi}(t),
\end{align*}
where $\Gamma_{(p+1)}$ is an $M_y \times M_x^{p}$ matrix obtained by matricizing a $(p+1)$-dimensional $M_x \times \cdots \times M_x \times M_y$ tensor $\underline{\Gamma} = (\gamma_{k_1,\ldots,k_{p+1}})_{k_1,\ldots,k_{p+1}}$ with respect to the $(p+1)$-th array.
Therefore, we can easily extend the quadratic model to the $p$-th order polynomial model and estimate it similarly using the method described in the next section, but in the following sections, we return to the functional quadratic model (\ref{eq:model}).  
\section{Model estimation and evaluation}
We consider estimating the model in a penalized likelihood framework of the penalized likelihood method.  
To do this, we assume that the error function $\varepsilon_i(t)$ has the following structure \citep{FaZh2000,ShCh2011}:
\begin{align}
	&\varepsilon_i(t) = \tau_i(t) + e_i(t), \label{eq:GP}\\
	\tau_i(t)\sim GP(0, k(\cdot, \cdot)), ~~
	&k(t, t') = \nu_1 \exp\left\{-\frac{\nu_2}{2}(t-t')^2\right\},~~
	e_i(t)\overset{\text\small\textrm{i.i.d.}}{\sim} N(0, \nu_3),
	\nonumber
\end{align}
where $GP(0, k(\cdot,\cdot))$ denotes a Gaussian process with mean 0 and a covariance function $k(\cdot, \cdot)$, and $\nu_1>0$, $\nu_2>0$, and $\nu_3>0$ are additional model parameters.  
By considering that the response $\bm y_i = (y_{i1},\ldots, y_{in_i})^T$ is observed at $n_i$ time points $t_{i1}, \ldots, t_{in_i}$ for each subject, we have the following probability density function:
\begin{align*}
	f(\bm y_i| \Theta, \bm{\nu}) = \frac{1}{(2\pi)^{n_i/2}\sqrt{|\Sigma_i|}}\exp\left\{
	-\frac{1}{2}\left(\bm y_i - \Psi_i\Theta \bm z_i \right)^T\Sigma_i^{-1}
	(\bm y_i - \Psi_i\Theta \bm z_i)
	\right\},
\end{align*}
where $\bm{\nu} = (\nu_1, \nu_2, \nu_3)^T$, $\Sigma_i = K_i + \nu_3^2I_{n_i}$, $K_i = (k(t_{ij}, t_{ij'}))_{jj'}$ and $\Psi_i = (\bm\psi(t_{i1}), \ldots, \bm\psi(t_{in_i}))^T$.
\subsection{Penalized likelihood method}
To estimate parameters $\Theta$ and $\bm{\nu}$, we consider maximizing the penalized log-likelihood function given by
\begin{align}
\ell_\lambda (\Theta,\bm{\nu}) = 
\ell (\Theta,\bm{\nu}) - \frac{n\lambda}{2}P(\Theta),
\label{eq:penlike}
\end{align}
where $\ell (\Theta,\bm{\nu}) = \sum_{i=1}^n \log f(\bm y_i| \Theta, \bm{\nu})$ is a log-likelihood function, $\lambda>0$ is a regularization parameter, and $P(\Theta)$ is a penalty function.  
To penalize the coefficient functions in the model (\ref{eq:model}) for the fluctuation in the $r$, $s$, and $t$ directions for linear and quadratic terms, we configure the following penalty function:
\begin{align}
P(\Theta) =& 
\bm{\alpha}^T\Omega_y \bm{\alpha} + 
{\rm tr}\left\{B^T\Omega_x B\right\} + 
{\rm tr}\left\{B\Omega_y B^T\right\} + \label{eq:pen}\\
& {\rm tr}\left\{\Gamma_{(3)}(\Omega_x\otimes I_{M_x}) \Gamma_{(3)}^T\right\} + 
{\rm tr}\left\{\Gamma_{(3)}(I_{M_x}\otimes \Omega_x) \Gamma_{(3)}^T\right\} + 
{\rm tr}\left\{\Gamma_{(3)}^T\Omega_y \Gamma_{(3)}\right\} \nonumber\\
=&
{\rm tr}\left\{\Theta^T\Omega_x^*\Theta\right\} + 
{\rm tr}\left\{\Theta\Omega_y\Theta^T\right\} \nonumber\\
=& ({\rm vec}\Theta)^T\Omega ({\rm vec}\Theta), \nonumber
\end{align}
where $\Omega_x$ and $\Omega_y$ are respectively $M_x \times M_x$ and $M_y \times M_y$ positive semi-definite matrices.  
An example for $\Omega_x$ and $\Omega_y$ is to use $D_2^TD_2$ with a second-order differential matrix $D_2$.  
Furthermore, $\Omega_x^* = {\rm blockdiag}\{0, \Omega_x, \Omega_x\otimes I_{M_x}+I_{M_x}\otimes \Omega_x\}$ and $\Omega = I_{M_y}\otimes \Omega_x^* + \Omega_y\otimes I_{1+M_x+M_x^2}$.  
The 1st term of the second equation of (\ref{eq:pen}) corresponds to the penalty for the roughness of $\alpha(t)$, whilst the 2nd and 3rd terms penalize the roughness of $\beta(s,t)$ with respect to $s$ and $t$ directions, respectively.  
Furthermore, the 4th, 5th, and 6th terms penalize the roughness of $\gamma(r,s,t)$ with respect to the $r$, $s$, and $t$ directions, respectively.   
Then if $\bm\nu$ were known, $\Theta$ would be estimated as
\begin{align}
	{\rm vec}\hat{\Theta} &= 
	\left\{
	\sum_{i=1}^n X_i^T\Sigma_i^{-1}X_i
	+ n\lambda\Omega\right\}^{-1}
	\left(\sum_{i=1}^n X_i^T\Sigma_i^{-1}\bm y_i \right)
	 \label{eq:Thetaest}\\
	&= \left(X^T\Sigma^{-1}X + n\lambda\Omega \right)^{-1}\left(X^T\Sigma^{-1}\bm y\right), \nonumber
\end{align}
where $X=(X_1^T,\ldots, X_n^T)^T$, $X_i = \bm z_i^T\otimes\Psi_i$, $\Sigma = {\rm blockdiag}\{\Sigma_1,\ldots, \Sigma_n\}$, and $\bm y = (\bm y_1^T,\ldots, \bm y_n^T)^T$.    
In practice ${\bm{\nu}}$ is unknown and it is difficult to derive the estimator of $\bm{\nu}$ analytically. 
Therefore, several iterative algorithms have been proposed and used in the literature. 
Here we apply the Newton-Raphson method, updated by
\begin{align}
	\bm\nu^{new} = \bm\nu^{old} - 
	\left\{\left.\frac{\partial^2 \ell_\lambda(\Theta, \bm{\nu})}{\partial \bm\nu\partial \bm\nu^T}\right|_{\bm{\nu}=\bm{\nu}^{old}}\right\}^{-1}
	\left.\frac{\partial \ell_\lambda(\Theta, \bm{\nu})}{\partial \bm\nu}\right|_{\bm{\nu}=\bm{\nu}^{old}}. \label{eq:NR}
\end{align}
The first and second derivatives of $\ell_\lambda(\Theta, \bm{\nu})$ with respect to $\nu_j$ $(j=1,2,3)$ are provided in the Appendix.
Parameters $\Theta$ and $\bm{\nu}$ are alternately updated until convergence, and then we arrive at penalized maximum likelihood estimators $\hat{\Theta}$ and $\hat{\bm{\nu}}$, respectively.  
Finally, we have a statistical model $f(\bm y_i|\hat{\Theta}, \hat{\bm{\nu}})$.  
\subsection{Model selection criteria}
\label{sec:IC}
It is important to select optimal values of tuning parameters such as the number of basis functions $M_y$ in (\ref{eq:basis_b}) and the regularization parameter $\lambda$ in (\ref{eq:penlike}).  
Here we introduce some model selection criteria for evaluating the estimated model.  

First, we introduce model selection criteria based on effective degrees of freedom described in \cite{EiMa1996}.  
To derive the effective degrees of freedom of our model, we express the predicted value of the response $\hat{\bm y}$ as
\begin{align*}
\hat{\bm y}
= X{\rm vec}{\hat{\Theta}} 
= X\left\{
X^T\Sigma^{-1}X + n\lambda\Omega\right\}^{-1}
X^T\Sigma^{-1}\bm y.  
\end{align*}
Using \cite{EiMa1996}, the effective degrees of freedom are $df = {\rm tr}\{S\}$ where $S = X\left\{
X^T\Sigma^{-1}X + n\lambda\Omega \right\}^{-1}X^T\Sigma^{-1}$ is known as a hat matrix or a smoother matrix.  
If $\lambda=0$ and $n > (1+M_x+M_x^2)M_y$ then the effective degrees of freedom becomes $(1+M_x+M_x^2)M_y$, which is just the same as the number of parameters included in $\Theta$.  
The GCV and the modified AIC are respectively given by
\begin{align*}
	{\rm GCV} &= \frac{1}{n}\frac{(\bm y - \hat{\bm y})^T(\bm y - \hat{\bm y})}{\displaystyle{\left\{1-(df+3) / n\right\}^2}}, \\
	{\rm mAIC} &= -2\ell(\hat\Theta, \hat{\bm{\nu}}) +2(df+3).
\end{align*}
Note that we added the number of parameters in $\bm{\nu}$, 3, to the degrees freedom.  

The AIC was originally derived for evaluating models estimated by the maximum likelihood method, so it is not suitable for evaluating those estimated by the penalized likelihood method. 
To solve this problem, \cite{KoKi2008} derived information criteria for evaluating models estimated by penalized likelihood methods.  
Specifically, rooted in information theory and a Bayesian approach, they proposed a generalized information criterion (GIC) and a generalized Bayesian information criterion (GBIC).  
Based on \cite{KoKi1996} and \cite{KoAnIm2004}, the GIC and GBIC are respectively given by
\begin{align}
	{\rm GIC} =& -2\ell(\hat\Theta, \hat{\bm{\nu}}) +2{\rm tr}\left\{R(\hat{\Theta},\hat{\bm\nu})^{-1}
	Q(\hat{\Theta},\hat{\bm\nu})\right\}. \nonumber\\
	{\rm GBIC} =& -2\ell(\hat\Theta, \hat{\bm{\nu}}) + 
	n\lambda({\rm vec}\hat{\Theta})^T\Omega {\rm vec}\hat{\Theta} -(\eta - \zeta)\log\lambda + \zeta\log n +  \label{eq:IC2}\\ 
	& \log|R(\hat{\Theta},\hat{\bm\nu})| - \log|\Omega|_+ - \zeta\log(2\pi),\nonumber
\end{align}
where $R(\hat{\Theta},\hat{\bm\nu})$ and $Q(\hat{\Theta},\hat{\bm\nu})$ are respectively given by
\begin{align*}
R(\hat{\Theta}, \hat{\bm{\nu}}) &= -\frac{1}{n}
\begin{pmatrix}
\displaystyle{\left.\frac{\partial^2 \ell_\lambda(\Theta, \bm{\nu})}{\partial({\rm vec}\Theta)\partial({\rm vec}\Theta)^T}\right|_{\Theta=\hat{\Theta},~\bm{\nu}=\hat{\bm{\nu}}}} & 
\displaystyle{\left.\frac{\partial^2 \ell_\lambda(\Theta, \bm{\nu})}{\partial({\rm vec}\Theta)\partial(\bm{\nu})^T}\right|_{\Theta=\hat{\Theta},~\bm{\nu}=\hat{\bm{\nu}}}} \\ 
\displaystyle{\left.\frac{\partial^2 \ell_\lambda(\Theta, \bm{\nu})}{\partial\bm{\nu}\partial({\rm vec}\Theta)^T}\right|_{\Theta=\hat{\Theta},~\bm{\nu}=\hat{\bm{\nu}}}} & 
\displaystyle{\left.\frac{\partial^2 \ell_\lambda(\Theta, \bm{\nu})}{\partial\bm{\nu}\partial\bm{\nu}^T}\right|_{\Theta=\hat{\Theta},~\bm{\nu}=\hat{\bm{\nu}}}}  
\end{pmatrix} \\
Q(\hat{\Theta}, \hat{\bm{\nu}}) &= \frac{1}{n}
\begin{pmatrix}
\displaystyle{\left.\frac{\partial \ell_\lambda(\Theta, \bm{\nu})}{\partial({\rm vec}\Theta)} \frac{\partial \ell(\Theta, \bm{\nu})}{\partial({\rm vec}\Theta)^T}\right|_{\Theta=\hat{\Theta},~\bm{\nu}=\hat{\bm{\nu}}}} & 
\displaystyle{\left.\frac{\partial \ell_\lambda(\Theta, \bm{\nu})}{\partial({\rm vec}\Theta)}\frac{\partial \ell(\Theta, \bm{\nu})}{\partial\bm{\nu}^T}\right|_{\Theta=\hat{\Theta},~\bm{\nu}=\hat{\bm{\nu}}}} \\ 
\displaystyle{\left.\frac{\partial \ell_\lambda(\Theta, \bm{\nu})}{\partial\bm{\nu}}\frac{\partial \ell(\Theta, \bm{\nu})}{\partial({\rm vec}\Theta)^T}\right|_{\Theta=\hat{\Theta},~\bm{\nu}=\hat{\bm{\nu}}}} & 
\displaystyle{\left.\frac{\partial \ell_\lambda(\Theta, \bm{\nu})}{\partial\bm{\nu}}\frac{\partial \ell(\Theta, \bm{\nu})}{\partial\bm{\nu}^T}\right|_{\Theta=\hat{\Theta},~\bm{\nu}=\hat{\bm{\nu}}}}  
\end{pmatrix} 
\end{align*}
Details concerning the $R(\hat{\Theta}, \hat{\bm{\nu}})$, and $Q(\hat{\Theta}, \hat{\bm{\nu}})$ matrices are provided in the Appendix.  
Furthermore, $\eta = (1+M_x+M_x^2)M_y$ is the size of matrix $\Omega$, $\zeta$ is the number of nonzero eigenvalues of $\Omega$, and $|\cdot|_+$ denotes the product of positive matrix eigenvalues.  
We select tuning parameter values that minimize these criteria and then treat the corresponding model as the optimal candidate.
\section{Simulations}
We conducted Monte Carlo simulations to investigate the efficacy of the proposed method.  
We artificially generated datasets from a regression model, and then compared the prediction accuracy of the proposed method with ordinary models.  

The data generation procedure is given as follows.  
First we assume that the relationship between the functional predictor and the functional response is given by the functional quadratic model (\ref{eq:model}) with basis expansions (\ref{eq:basis_x}) and (\ref{eq:basis_b}). 
We set $\alpha(t)=0$ for simplicity.  
We used $B$-splines for basis functions $\bm{\phi}(s)$ and $\bm{\psi}(t)$, where the numbers of these basis functions, respectively $M_x$ and $M_y$, are fixed at seven.  
The elements of matrices $B$ and $\Gamma$ included in (\ref{eq:basis_b}) are generated from a Wishart distribution with 10 degrees of freedom and a Toeplitz scale matrix.  
Similarly, to generate the predictor function $x_i(t)$, the coefficients $\bm w_i$ in (\ref{eq:basis_x}) are generated from a multivariate normal distribution with 0 mean vector and a Toeplitz variance covariance matrix.  
Then, the mean structure of the right-hand side of the model (\ref{eq:model}), denoted by $g_i(t)$ $(i=1,\ldots, n)$, is obtained.  
There are $n_t = 21$ equally spaced time points.  
We also added noise generated using a Gaussian process given in (\ref{eq:GP}) to $g_i(t)$ and then configured a longitudinal response $y_{ij}$, where subscript $j$ indexes time.  
We then added Gaussian noise to the predictor $x_i(t)$ at discrete time points and generated longitudinal data $x_{ij}$.  
We treated $x_{ij}$ and $y_{ij}$ as observations, and therefore the objective of the simulation is to predict $g_i(t)$ using $x_{ij}$ and $y_{ij}$.   

Next, we analyzed data by the following procedure.  
We fitted $x_{ij}$ for each $i$ to construct functional data $x_i(t)$ using a smoothing technique with Gaussian radial basis functions \citep{KaKo2007}. 
To avoid computational burden, we fixed the number $M_y$ of basis functions at six.  
After obtaining $\bm w_i$, we estimated the model using the proposed method.  
Here we conduct a comparative exercise to evaluate the proposed method.  

\begin{figure}[t]
	\begin{center}
		\begin{tabular}{c}
			\begin{minipage}{0.33\hsize}
				\begin{center}
					\includegraphics[width=5cm]{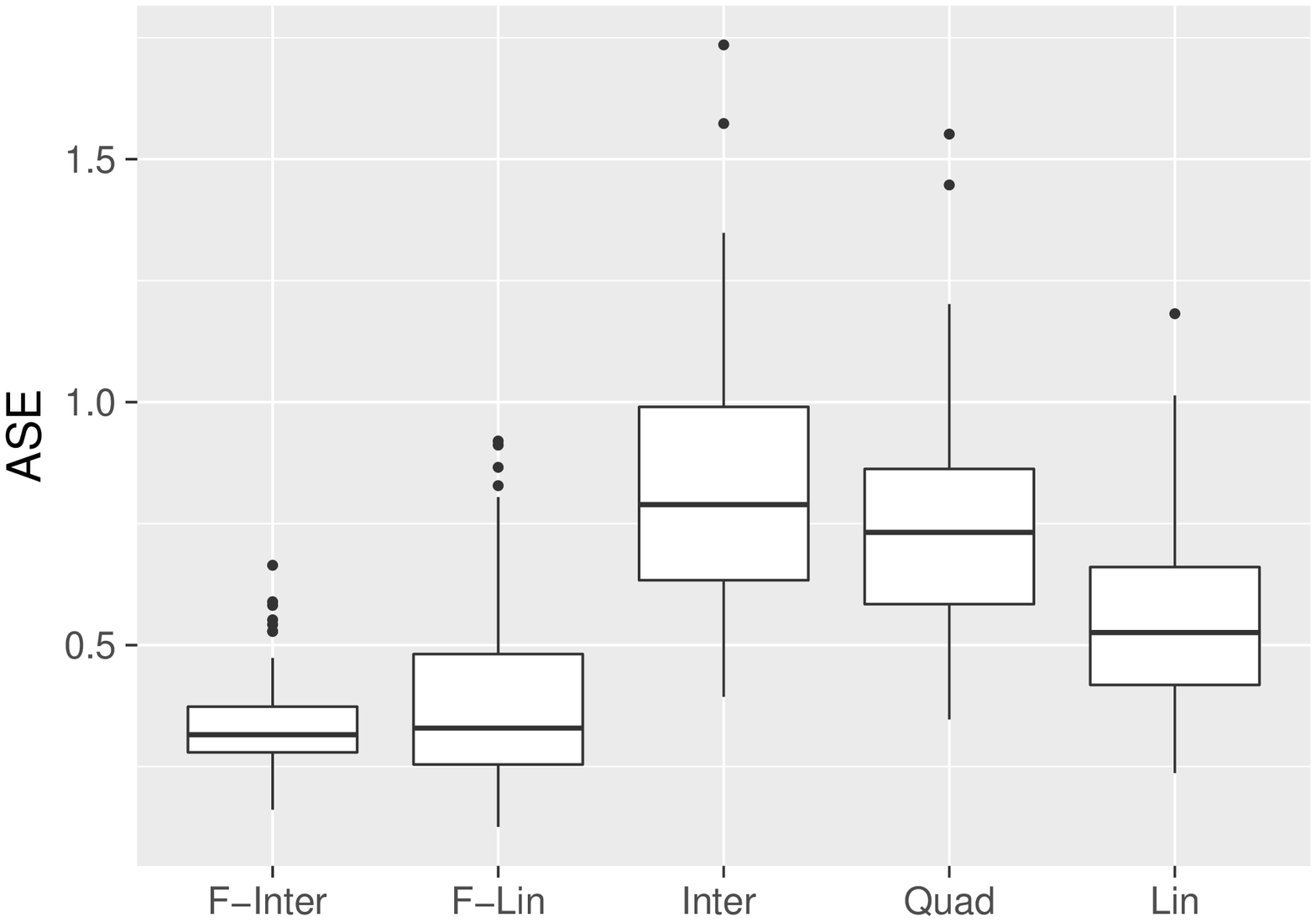} 
					{\scriptsize $n=50,~\nu_3=0.3$}
				\end{center}
			\end{minipage}
			\begin{minipage}{0.33\hsize}
				\begin{center}
					\includegraphics[width=5cm]{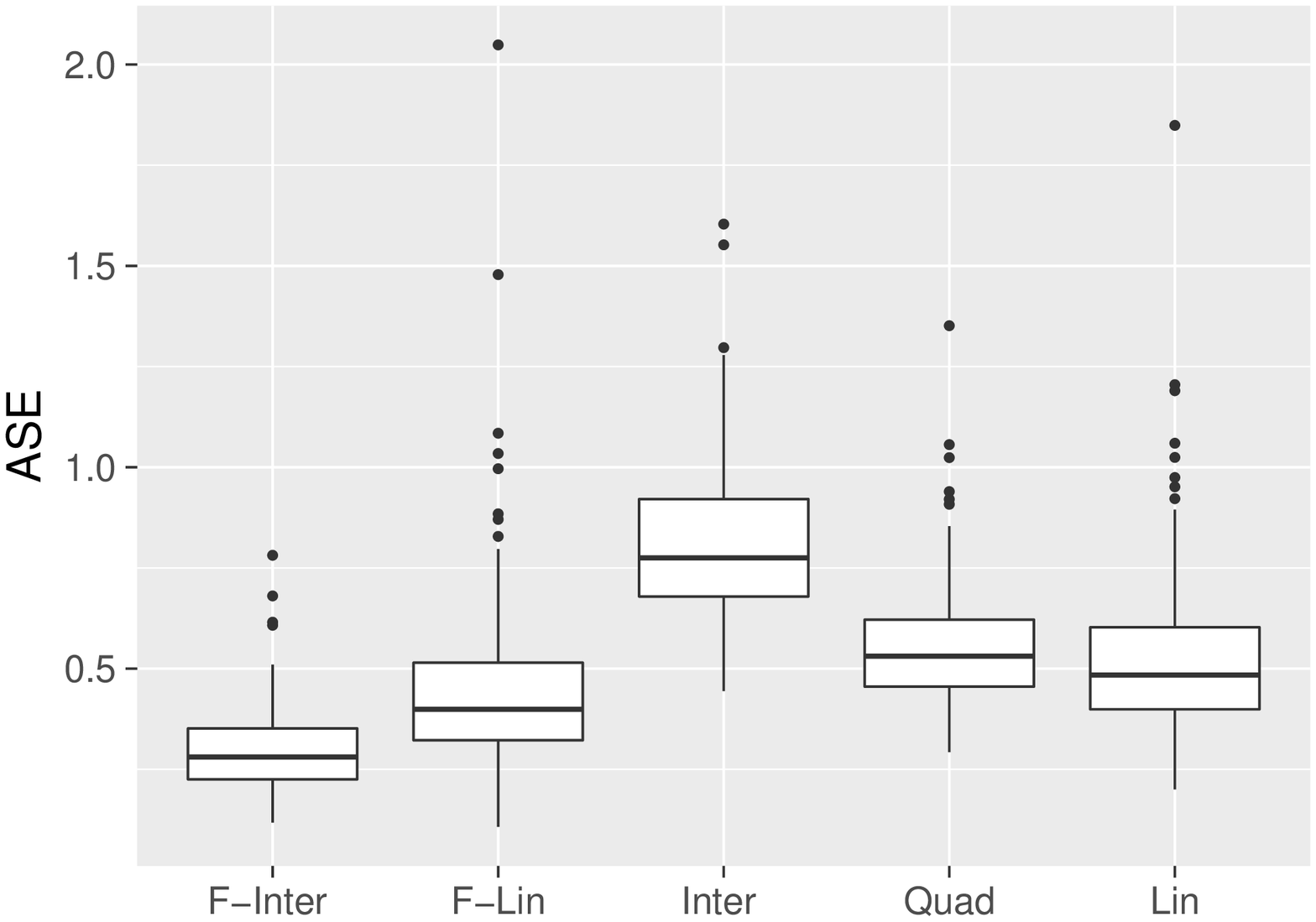} 
					{\scriptsize $n=100,~\nu_3=0.3$}
				\end{center}
			\end{minipage}
			\begin{minipage}{0.33\hsize}
				\begin{center}
					\includegraphics[width=5cm]{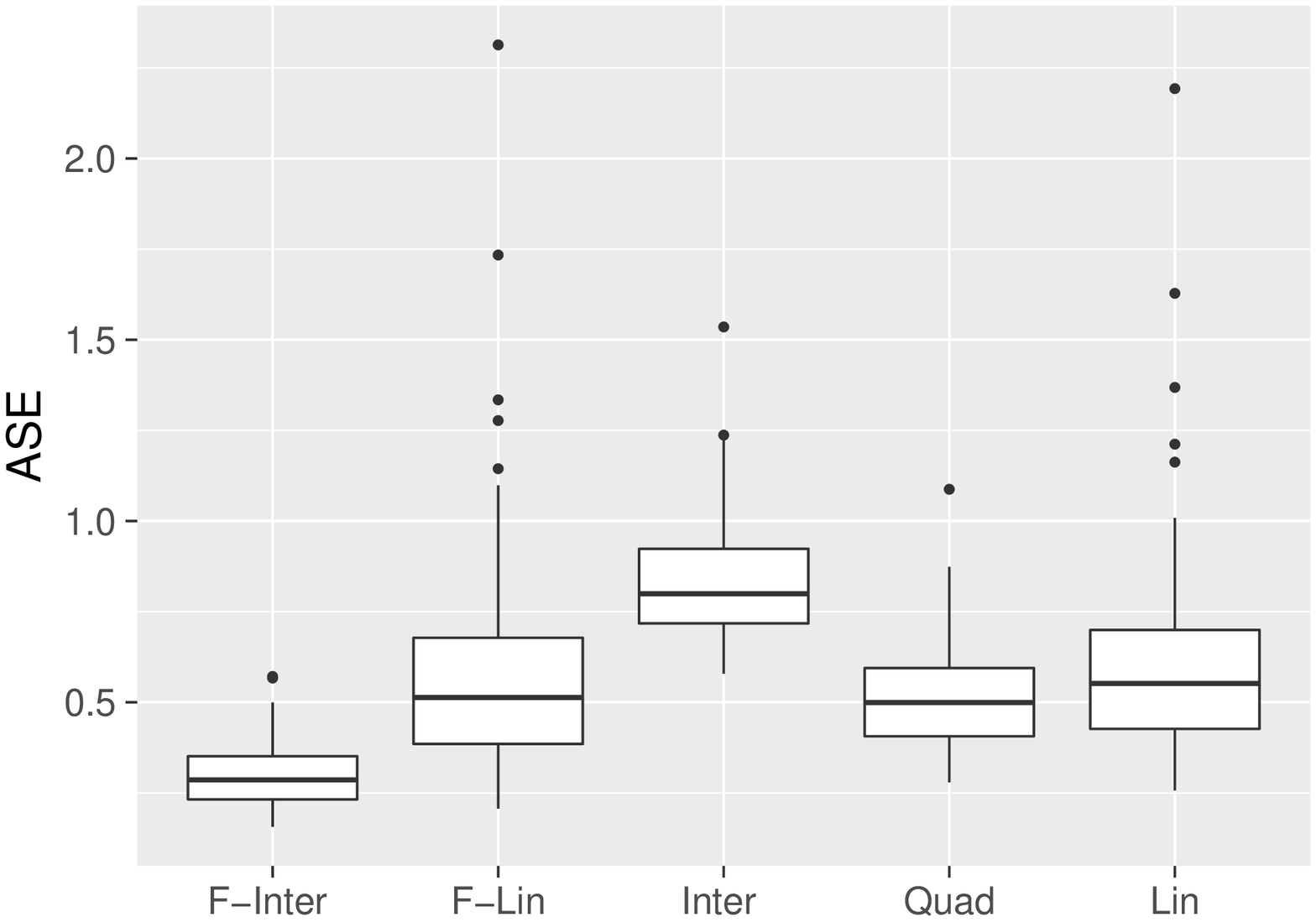} 
					{\scriptsize $n=200,~\nu_3=0.3$}
				\end{center}
			\end{minipage}
		\end{tabular}
		\begin{tabular}{c}
			\begin{minipage}{0.33\hsize}
				\begin{center}
					\includegraphics[width=5cm]{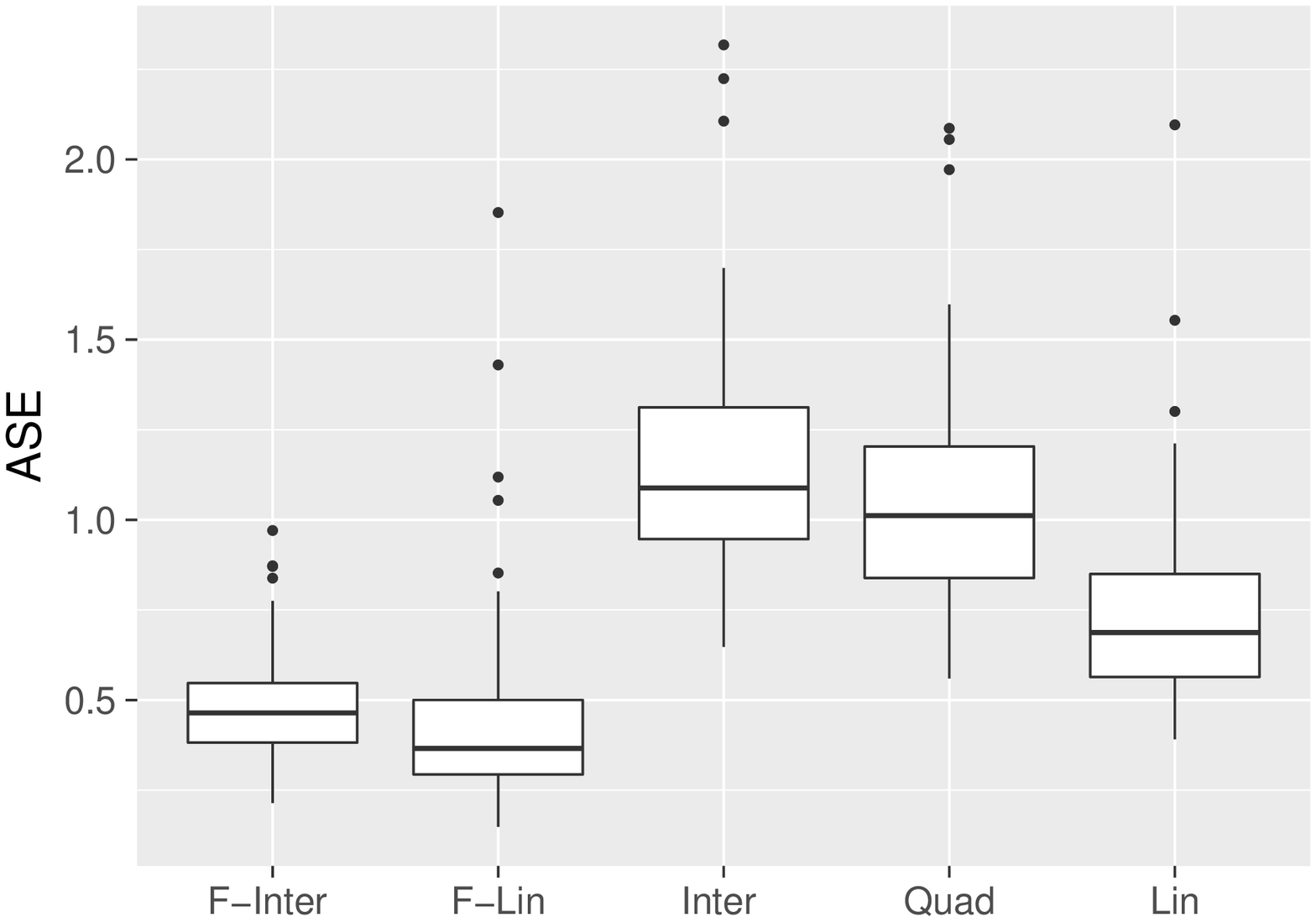} 
					{\scriptsize $n=50,~\nu_3=0.6$}
				\end{center}
			\end{minipage}
			\begin{minipage}{0.33\hsize}
				\begin{center}
					\includegraphics[width=5cm]{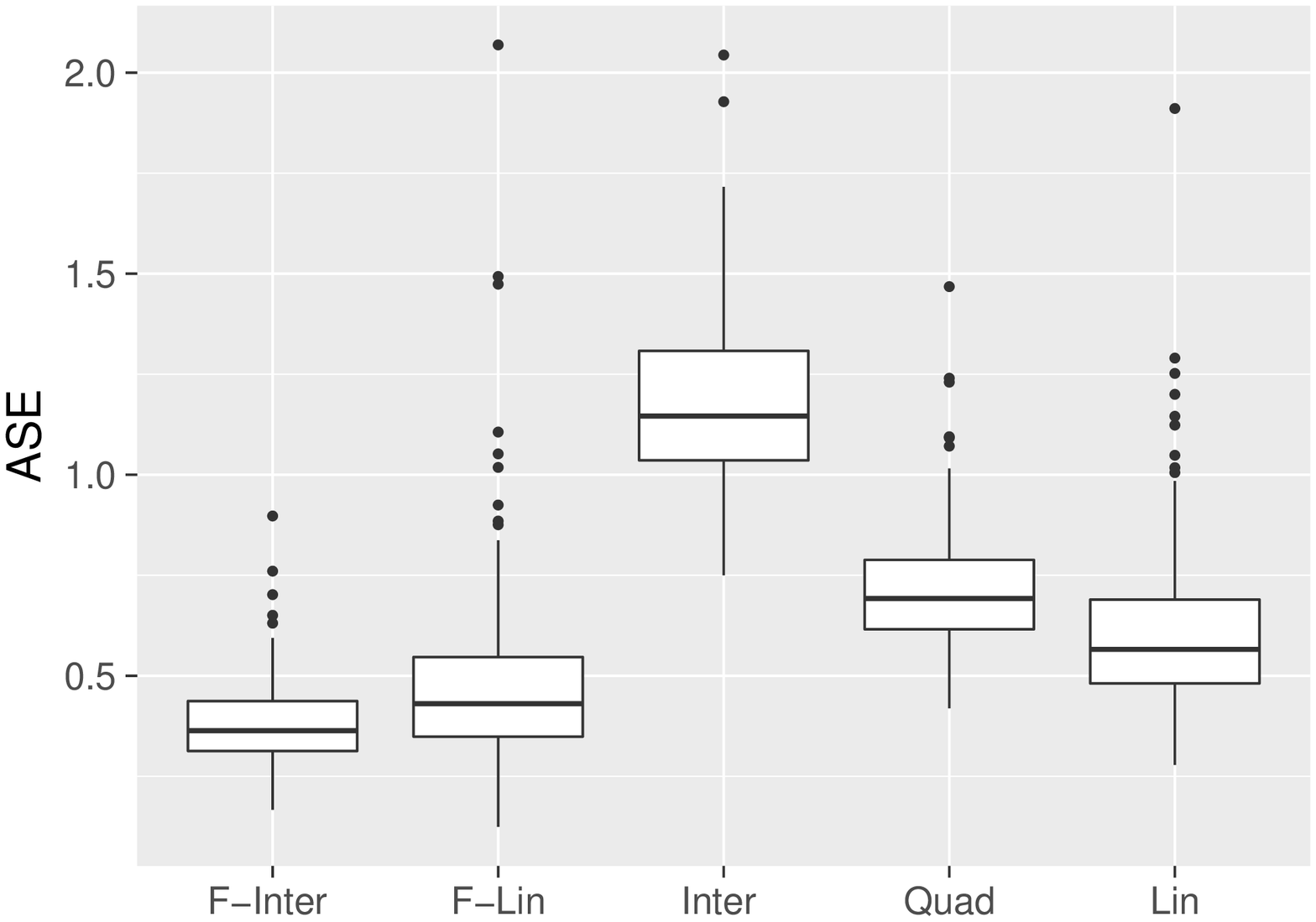} 
					{\scriptsize $n=100,~\nu_3=0.6$}
				\end{center}
			\end{minipage}
			\begin{minipage}{0.33\hsize}
				\begin{center}
					\includegraphics[width=5cm]{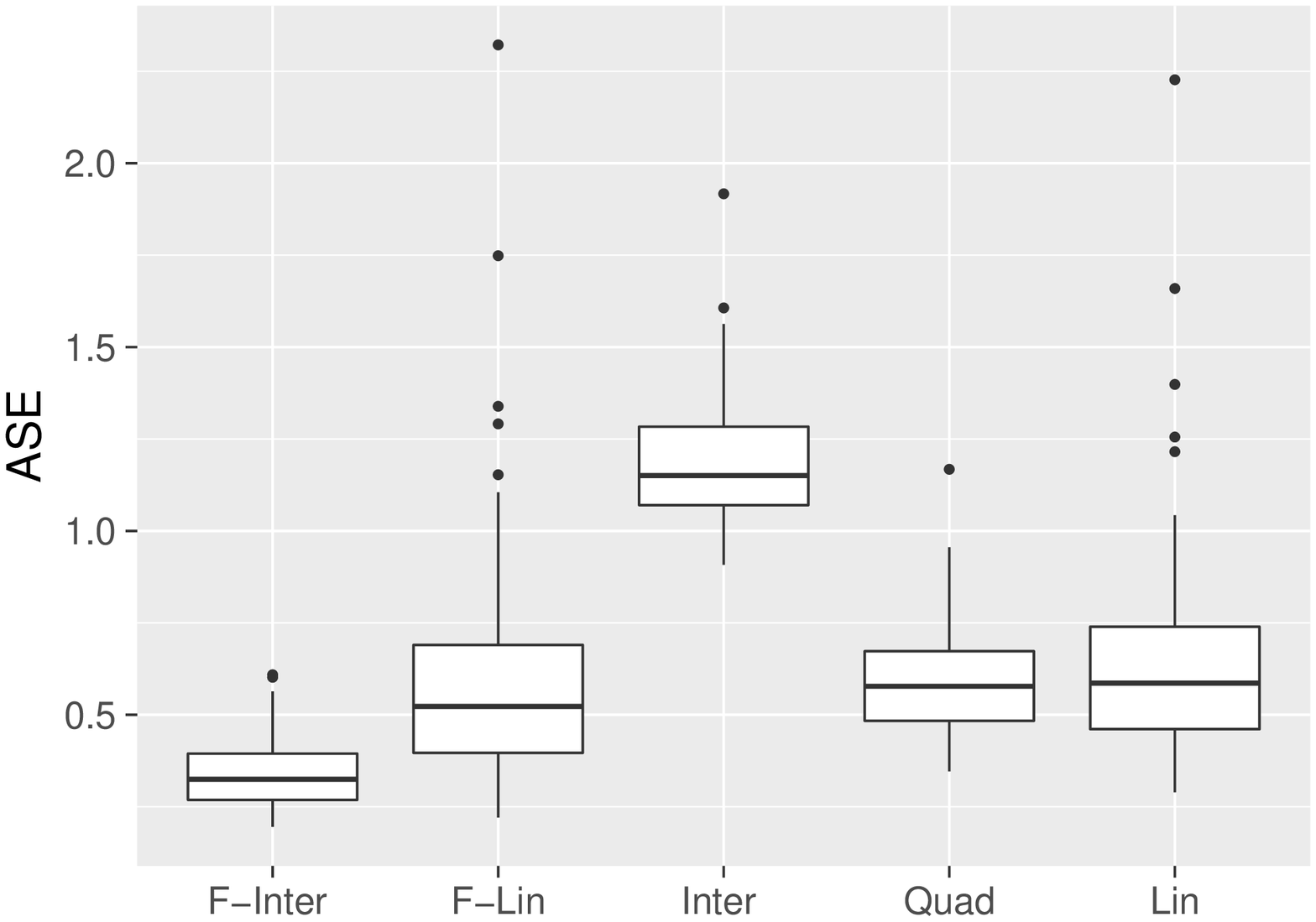} 
					{\scriptsize $n=200,~\nu_3=0.6$}
				\end{center}
			\end{minipage}
		\end{tabular}
	
		\caption[]{Boxplots for 100 ASEs obtained by the maximum likelihood estimation framework.  
		Each plot shows results of the proposed model (F-Quad.), functional linear model (F-Lin.), quadratic model (Quad.), and linear model (Lin.).}
		\label{fig:box1}
	\end{center}
%
	\begin{center}
		\begin{tabular}{c}
			\begin{minipage}{0.33\hsize}
				\begin{center}
					\includegraphics[width=5cm]{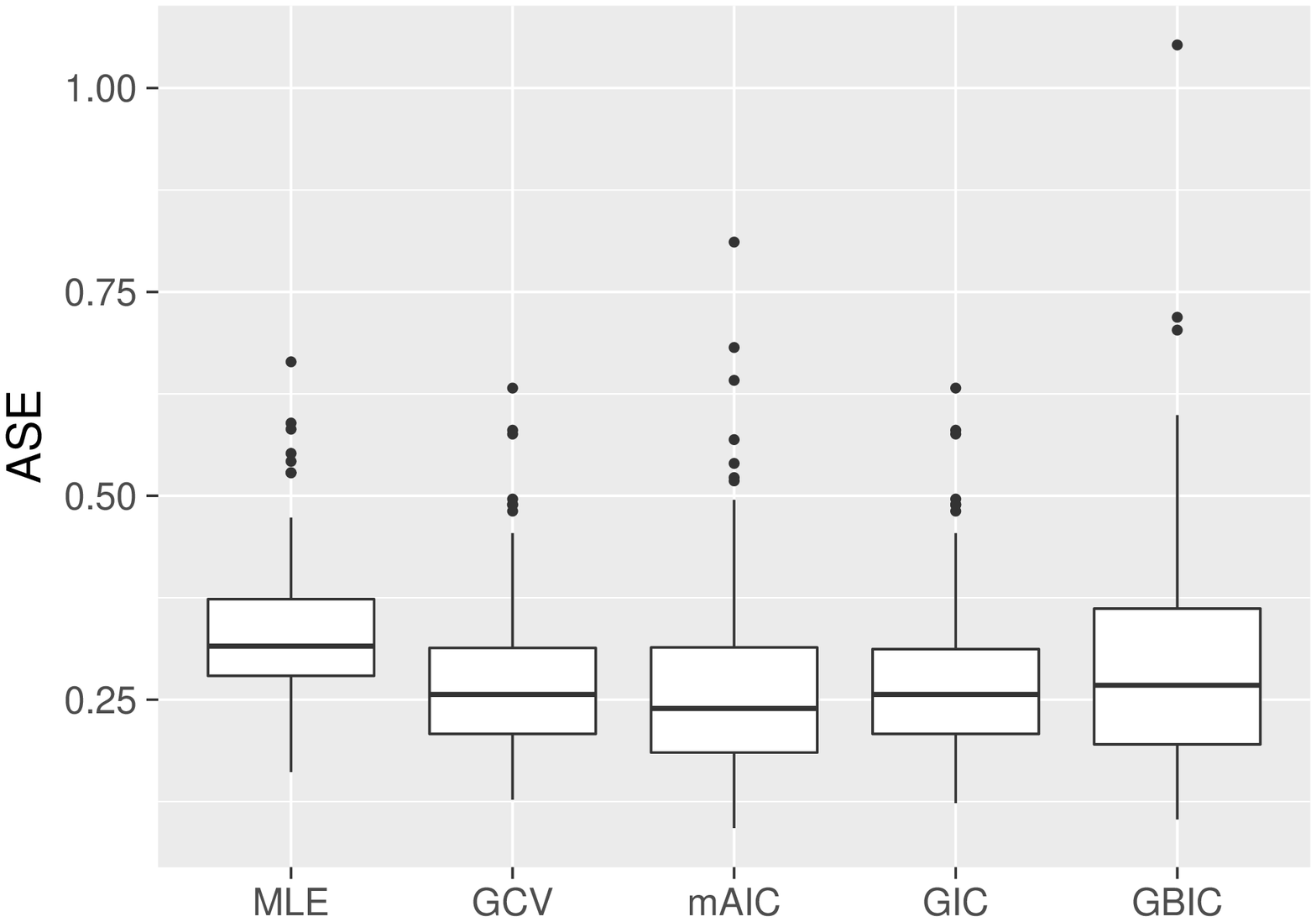} 
					{\scriptsize $n=50,~\nu_3=0.3$}
				\end{center}
			\end{minipage}
			\begin{minipage}{0.33\hsize}
				\begin{center}
					\includegraphics[width=5cm]{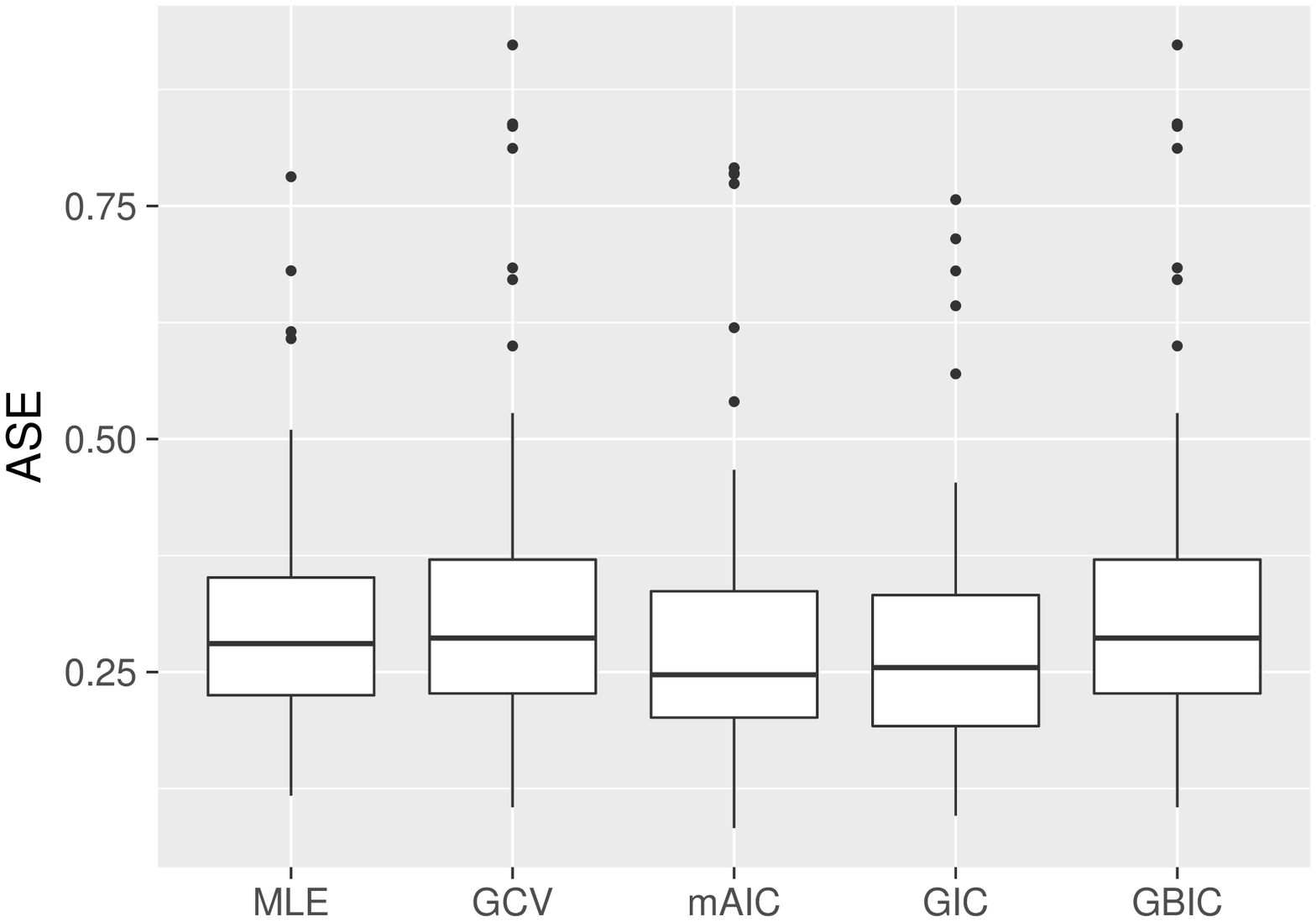} 
					{\scriptsize $n=100,~\nu_3=0.3$}
				\end{center}
			\end{minipage}
			\begin{minipage}{0.33\hsize}
				\begin{center}
					\includegraphics[width=5cm]{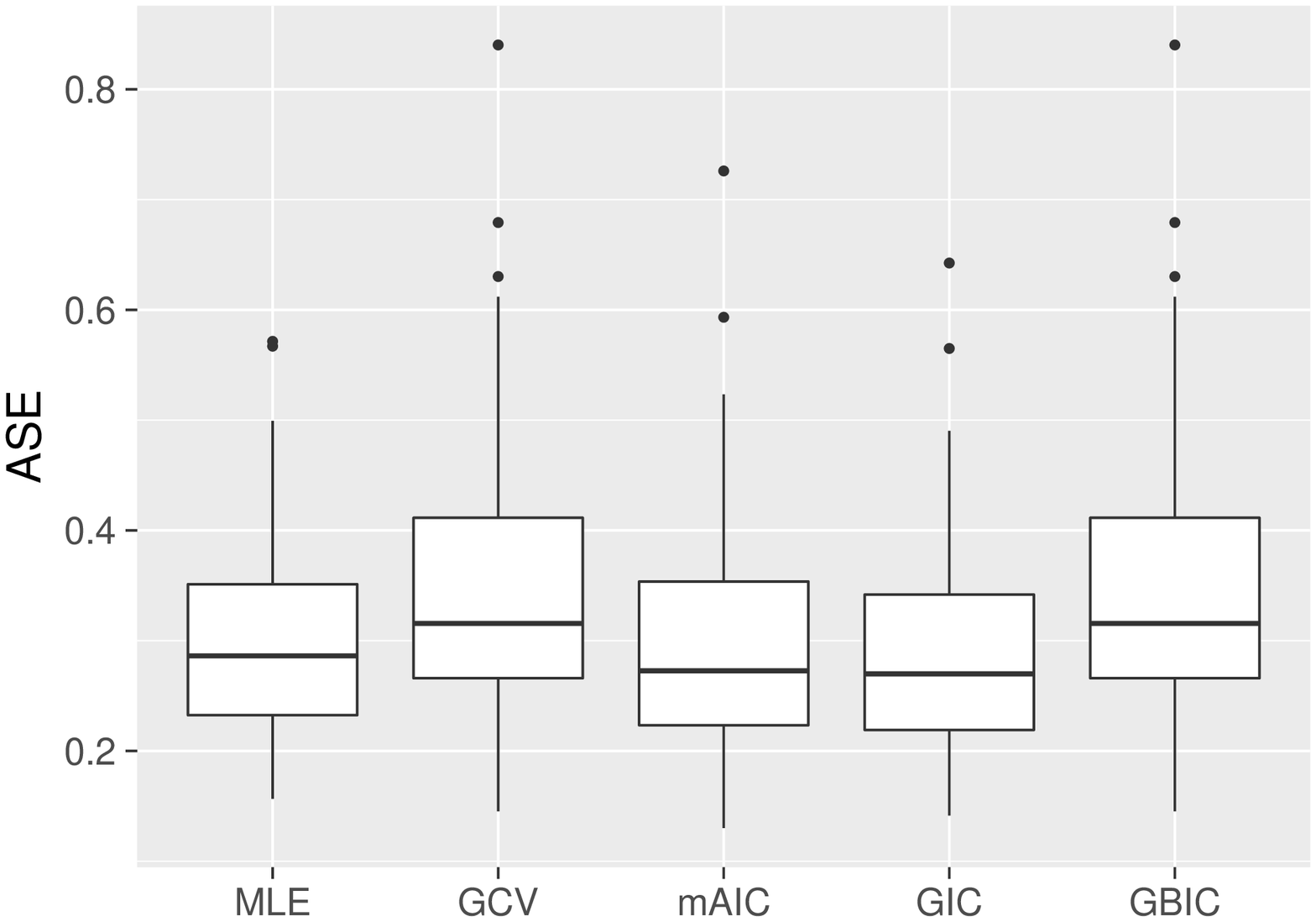} 
					{\scriptsize $n=200,~\nu_3=0.3$}
				\end{center}
			\end{minipage}
		\end{tabular}
		\begin{tabular}{c}
			\begin{minipage}{0.33\hsize}
				\begin{center}
					\includegraphics[width=5cm]{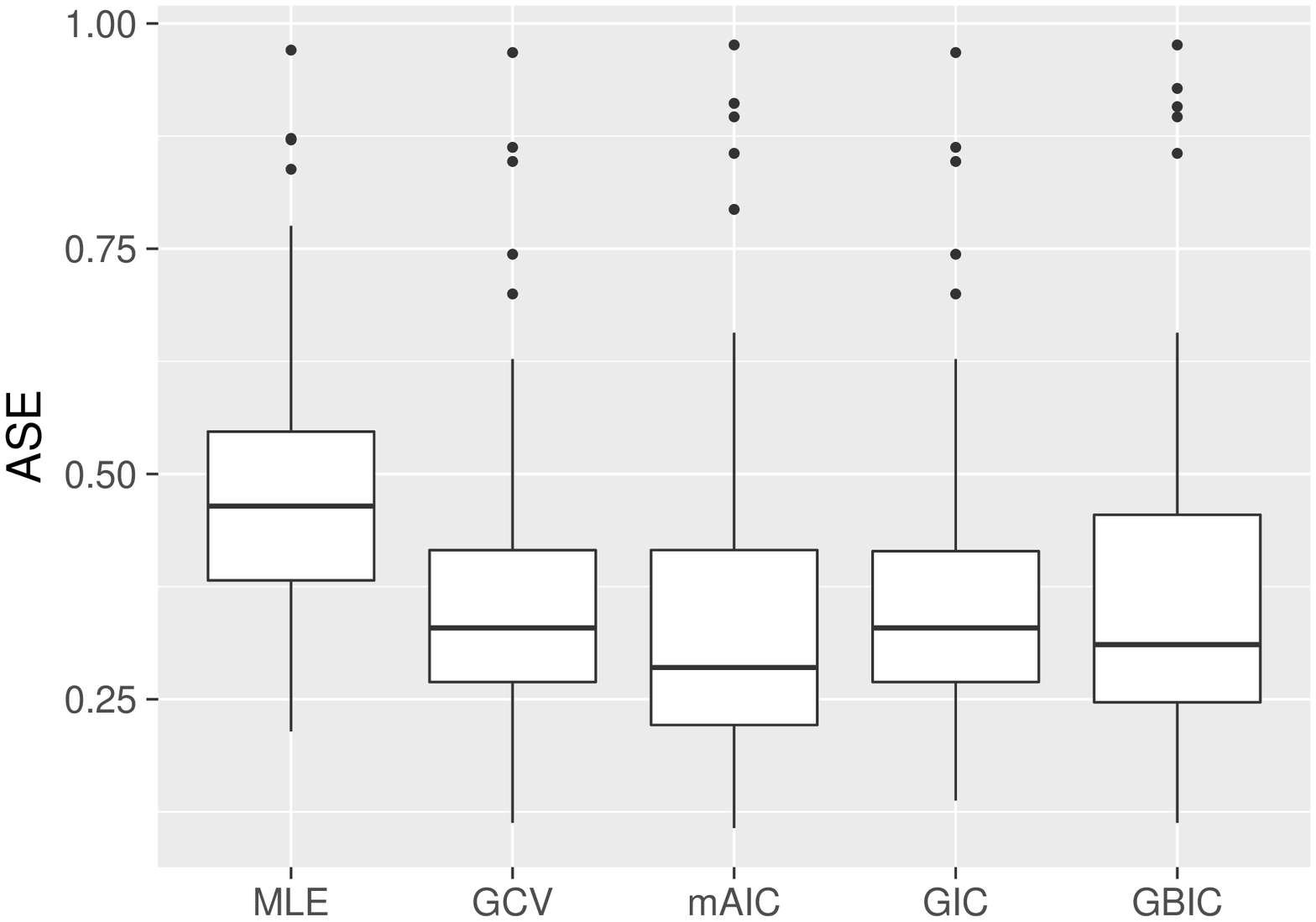} 
					{\scriptsize $n=50,~\nu_3=0.6$}
				\end{center}
			\end{minipage}
			\begin{minipage}{0.33\hsize}
				\begin{center}
					\includegraphics[width=5cm]{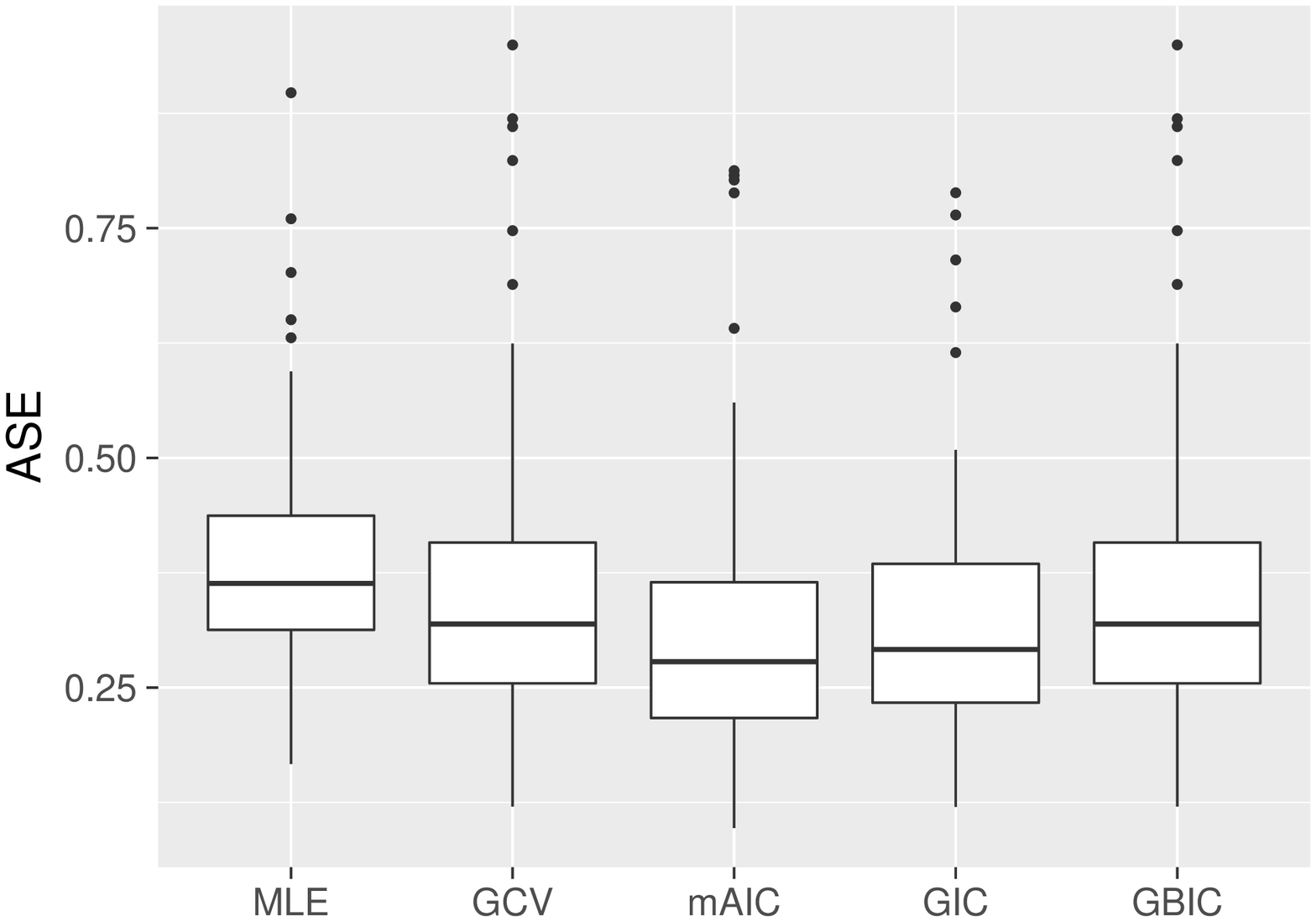} 
					{\scriptsize $n=100,~\nu_3=0.6$}
				\end{center}
			\end{minipage}
			\begin{minipage}{0.33\hsize}
				\begin{center}
					\includegraphics[width=5cm]{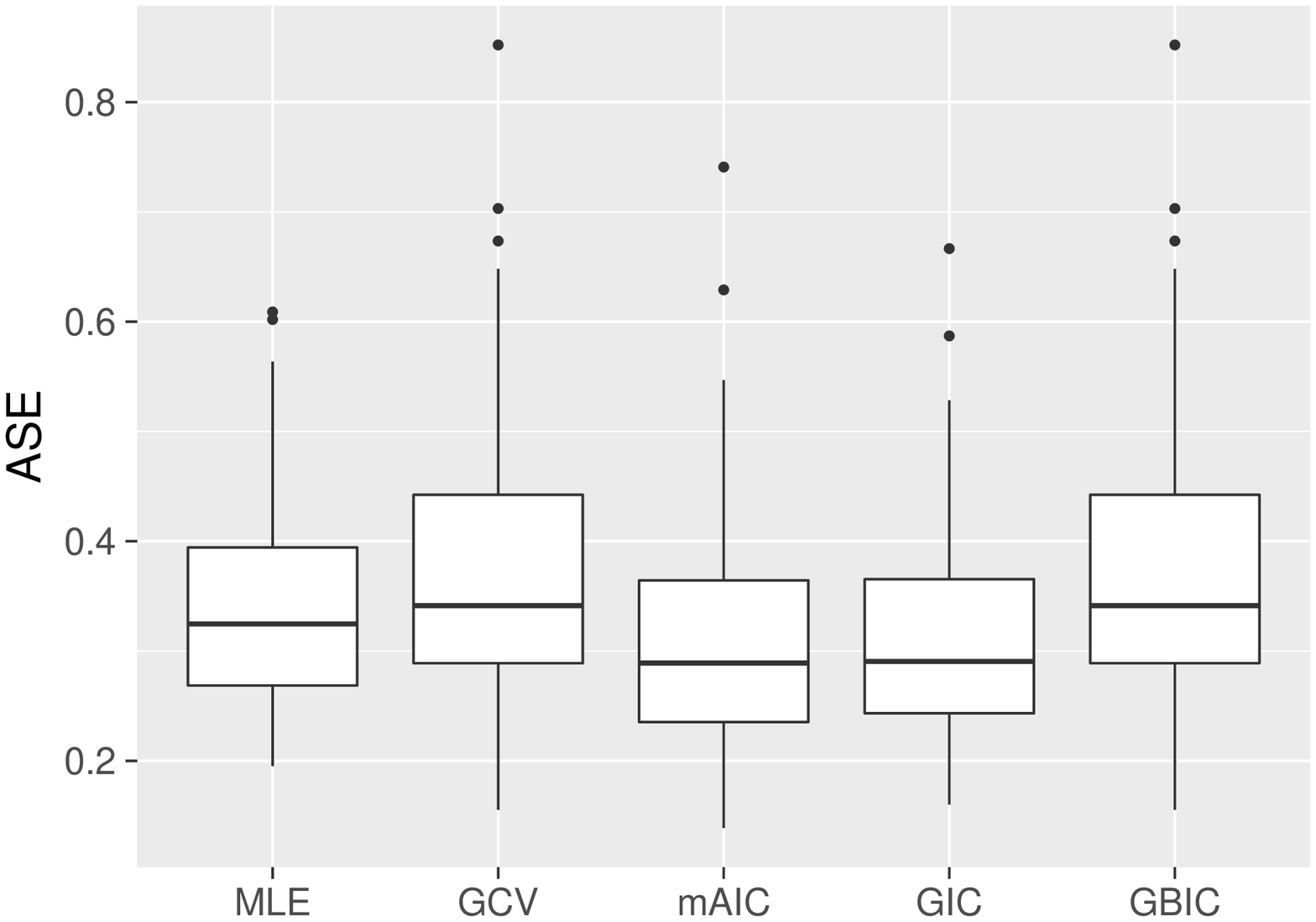} 
					{\scriptsize $n=200,~\nu_3=0.6$}
				\end{center}
			\end{minipage}
		\end{tabular}
		
		\caption[]{Boxplots for 100 ASEs obtained by penalized likelihood estimation. 
		Each plot shows the result of maximum likelihood estimation (MLE) and the GCV, mAIC, GIC, and GBIC model selection criterion.}
		\label{fig:box2}
	\end{center}
\end{figure}

For this exercise, first, we compared the proposed functional quadratic model with an interaction (F-INTER) with a functional linear model (F-LIN), a multivariate quadratic model with interactions (INTER), a quadratic model (QUAD), and a linear model (LIN).  
They are respectively given by
\begin{align*}
({\rm F-INTER})~~ & 	y_i(t)= \int x_{i}(s)\beta(s,t)ds + {\iint x_{i}(r)x_i(s)\gamma(r,s,t)drds} + \varepsilon_i(t), \\
({\rm F-LIN})~~ & y_i(t)= \int x_{i}(s)\beta(s,t)ds + \varepsilon_i(t),\\
({\rm INTER})~~ & y_{ij} = \sum_{j=1}^{n_t}x_{ij}\beta_j + \sum_{j,j'} x_{ij}x_{ij'}\gamma_{jj'} + \varepsilon_i(t), \\
({\rm QUAD})~~ & y_{ij} = \sum_{j=1}^{n_t}x_{ij}\beta_j + \sum_{j=1}^{n_t} x_{ij}^2\gamma_{j} + \varepsilon_i(t), \\
({\rm LIN})~~ & y_{ij} = \sum_{j=1}^{n_t}x_{ij}\beta_j + \varepsilon_i(t),  
\end{align*}
where $\beta_j$, $\gamma_{jj'}$, and $\gamma_{j}$ are coefficient parameters.  
We estimated the above five models using maximum likelihood estimation (MLE), but the generalized inverse is used for the inversion of $X^T\Sigma^{-1}X+n\lambda\Omega$ in (\ref{eq:Thetaest}).  
We used the generalized inverse because we wanted to compare results for small sample sizes, and in this case, the number of parameters exceeds the sample size for the proposed model and thus ordinal matrix inversion is not possible.  
For each model, we obtained the predicted response $\hat y_i(t)$ and then calculated the average squared error ${\rm ASE}= 1/(n\cdot n_t)\sum_{i=1}^n\sum_{j=1}^{n_t} (g_i(t_j)-\hat{y}_i(t_j))^2$.  
We repeated this procedure 100 times for several sample sizes and noise levels, and then investigated the prediction accuracy.  
Table \ref{tab:PMLE} presents averages and standard deviations for 100 ASEs across the five models, with boxplots of ASEs provided in \ref{fig:box1}.  
In most cases, the proposed model (F-INTER) minimizes ASE compared to other models, and it also provides the most stable results.  
F-LIN is associated with smaller ASEs than F-INTER for smaller sample sizes and larger noise levels, but it gives less stable results than F-INTER and its performance deteriorates as sample size increases.
ASEs for INTER are larger than other models, and have high variances, especially where the sample size is small.  
ASEs for QUAD are also large for small sample sizes, but they decrease as sample size increases.  
LIN gives smaller ASEs than other models for multivariate data, but they are still larger than the functional regression models.  

In the second analysis, penalized maximum likelihood estimation (PMLE) is used, and the optimal value of the regularization parameter $\lambda$ is decided by the four model selection criteria introduced in Section \ref{sec:IC}.  
As per the previous analysis, we calculated the ASE with 100 repetitions.  
Table \ref{tab:PMLE} and Figure \ref{fig:box2} show results according to the four model selection criteria.  
In all cases the penalized likelihood method yields better or competitive results than the maximum likelihood method, especially for smaller sample sizes.  
There is no remarkable difference among the results from the four model selection criteria, but the ASEs for GIC are more stable than for other criteria.  
\begin{table}[t]
	\centering
	\caption{Averages of 100 ASEs and their standard deviations (in parentheses) for five models estimated using MLE.  
	Values in this table are multiplied by 10.}
	\label{tab:MLE}
	\begin{tabular}{rrrrrrr}
		\hline
		& \multicolumn{2}{c}{$n=50$} & \multicolumn{2}{c}{$n=100$} & \multicolumn{2}{c}{$n=200$} \\
		& $\nu_3=0.3$ & $\nu_3=0.6$ & $\nu_3=0.3$ & $\nu_3=0.6$ & $\nu_3=0.3$ & $\nu_3=0.6$ \\ 
		\hline
		F-Inter & 3.35 (0.90) & 4.88 (1.41) & 3.05 (1.14) & 3.86 (1.21) & 3.02 (0.93) & 3.43 (0.95) \\ 
		F-Lin & 3.87 (1.85) & 4.34 (2.52) & 4.66 (2.73) & 5.00 (2.89) & 5.82 (3.08) & 5.93 (3.08) \\ 
		Inter & 8.16 (2.54) & 11.55 (3.14) & 8.16 (2.09) & 11.76 (2.26) & 8.33 (1.63) & 11.92 (1.69) \\ 
		Quad & 7.43 (2.32) & 10.41 (2.92) & 5.62 (1.72) & 7.20 (1.76) & 5.18 (1.38) & 5.92 (1.39) \\ 
		Lin & 5.60 (1.89) & 7.28 (2.53) & 5.40 (2.41) & 6.27 (2.46) & 6.10 (2.88) & 6.48 (2.87) \\ 
		\hline
	\end{tabular}
\end{table}
\begin{table}[t]
	\centering
	\caption{Averages of 100 ASEs and their standard deviations (in parentheses) for PMLE and model selection criteria.  
	Values in this table are multiplied by 10.}
	\label{tab:PMLE}
	\begin{tabular}{rrrrrrr}
		\hline
		& \multicolumn{2}{c}{$n=50$} & \multicolumn{2}{c}{$n=100$} & \multicolumn{2}{c}{$n=200$} \\
		& $\nu_3=0.3$ & $\nu_3=0.6$ & $\nu_3=0.3$ & $\nu_3=0.6$ & $\nu_3=0.3$ & $\nu_3=0.6$ \\ 
		\hline
		GCV & 2.77 (1.02) & 3.61 (1.51) & 3.28 (1.58) & 3.59 (1.58) & 3.49 (1.22) & 3.73 (1.23) \\ 
		mAIC & 2.69 (1.33) & 3.32 (1.76) & 2.85 (1.43) & 3.10 (1.44) & 2.98 (1.08) & 3.12 (1.10) \\ 
		GIC & 2.76 (1.02) & 3.61 (1.51) & 2.78  (1.24) & 3.21  (1.26) & 2.92  (1.00) & 3.13  (1.02) \\ 
		GBIC & 3.02  (1.58) & 3.61 (1.80) & 3.28  (1.58) & 3.59  (1.58) & 3.49  (1.22) & 3.73  (1.23) \\ 
		\hline
	\end{tabular}
\end{table}
\section{Empirical example}
\begin{figure}[t]
	\begin{center}
		\includegraphics[width=7cm]{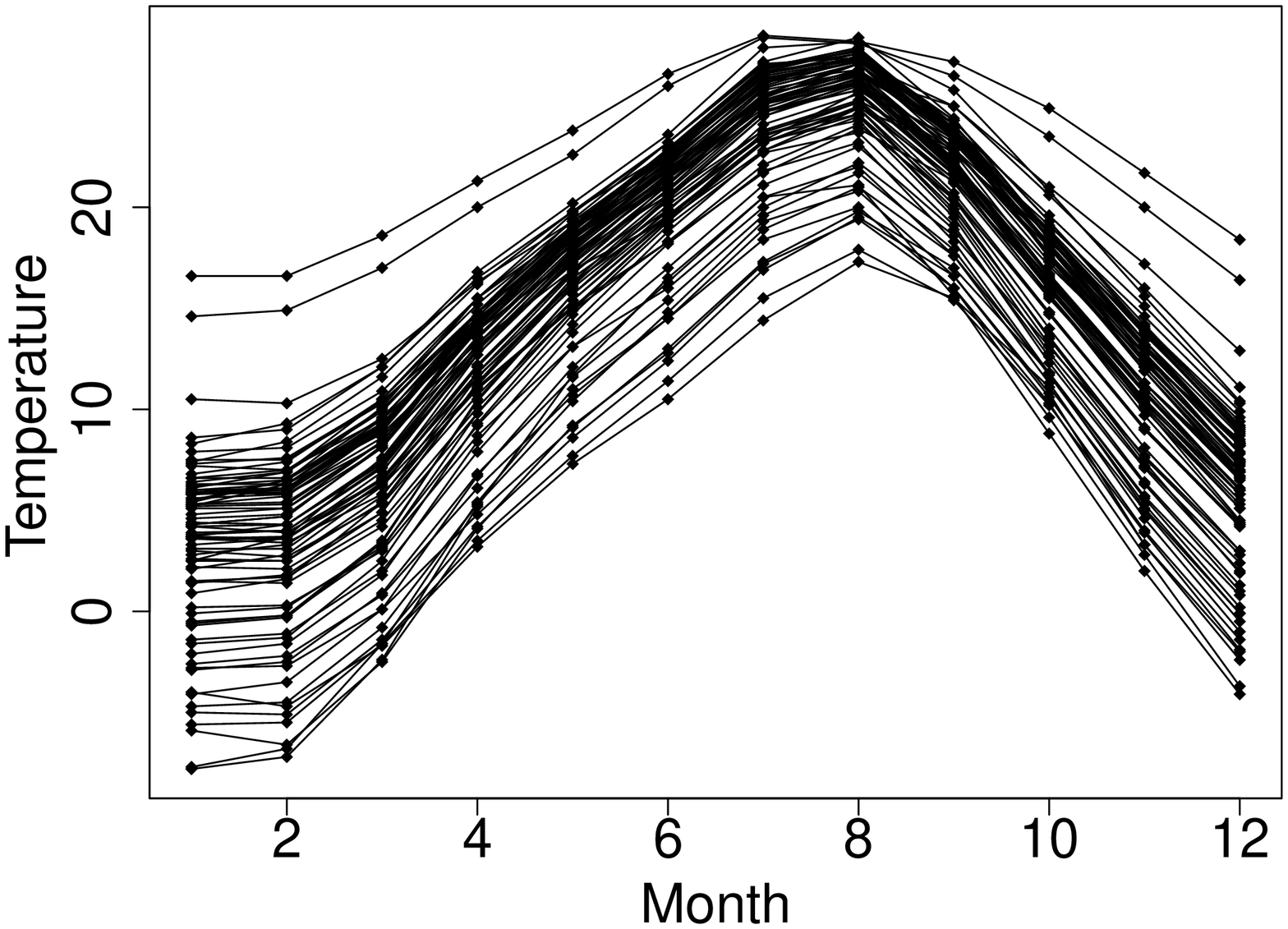} 
		\includegraphics[width=7cm]{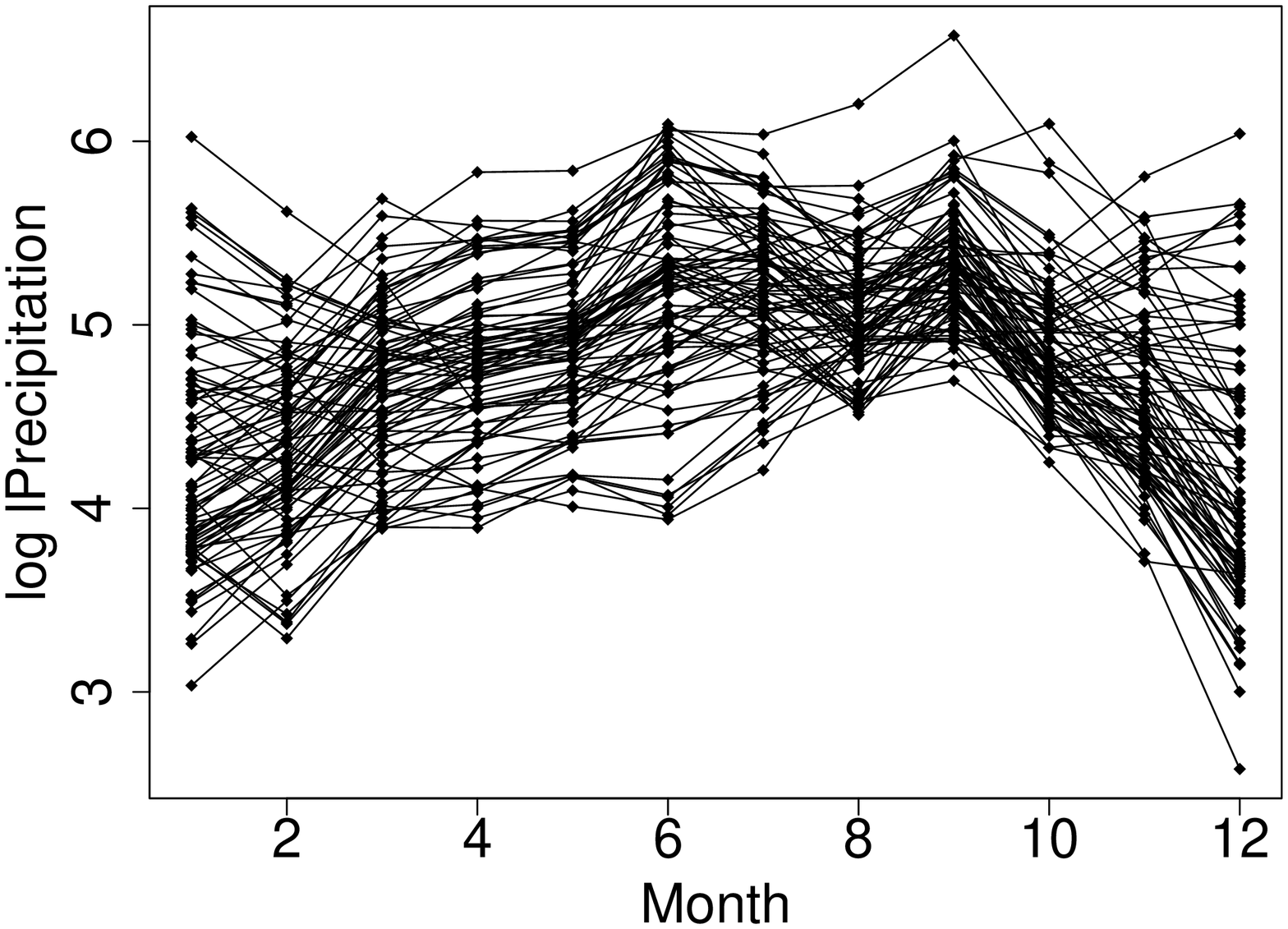}		
	\end{center}
	\caption[]{Weather data observed at cities in Japan.  
		Dots are observed data and each individual is connected by lines.  
		Left: monthly average temperature. Right: monthly total precipitation.
		}
	\label{fig:weatherdata}
\end{figure}

We applied the proposed method to the analysis of meteorological data available from Chronological Scientific Tables 2005.  
We used monthly temperature and the natural logarithm of monthly precipitation averaged from 1971 to 2000, observed at 76 cities in Japan.
The data are shown in Figure \ref{fig:weatherdata}.  
In many cities in Japan, it rains in summer more than in winter, especially in June. 
On the other hand, several cities have much snow in winter, and therefore these cities have much precipitation in winter.  
We treated temperature and precipitation as predictor and response, respectively, and considered data at 12 time points.   
%

First, we smoothed the temperature data using radial basis functions and then obtained coefficients $\bm w_i$ for functional data.
Next, we applied a functional quadratic model (\ref{eq:model}) to the data and estimated parameters $\Theta$ and $\bm{\nu}$ by the penalized likelihood method.  
The number of basis functions $M_y$ and the regularization parameter $\lambda$ are selected by model selection criterion GIC.  
Then we investigated the predicted precipitation and coefficient functions for linear and quadratic terms. 

\begin{figure}[t]
	\begin{center}
			\includegraphics[width=7cm]{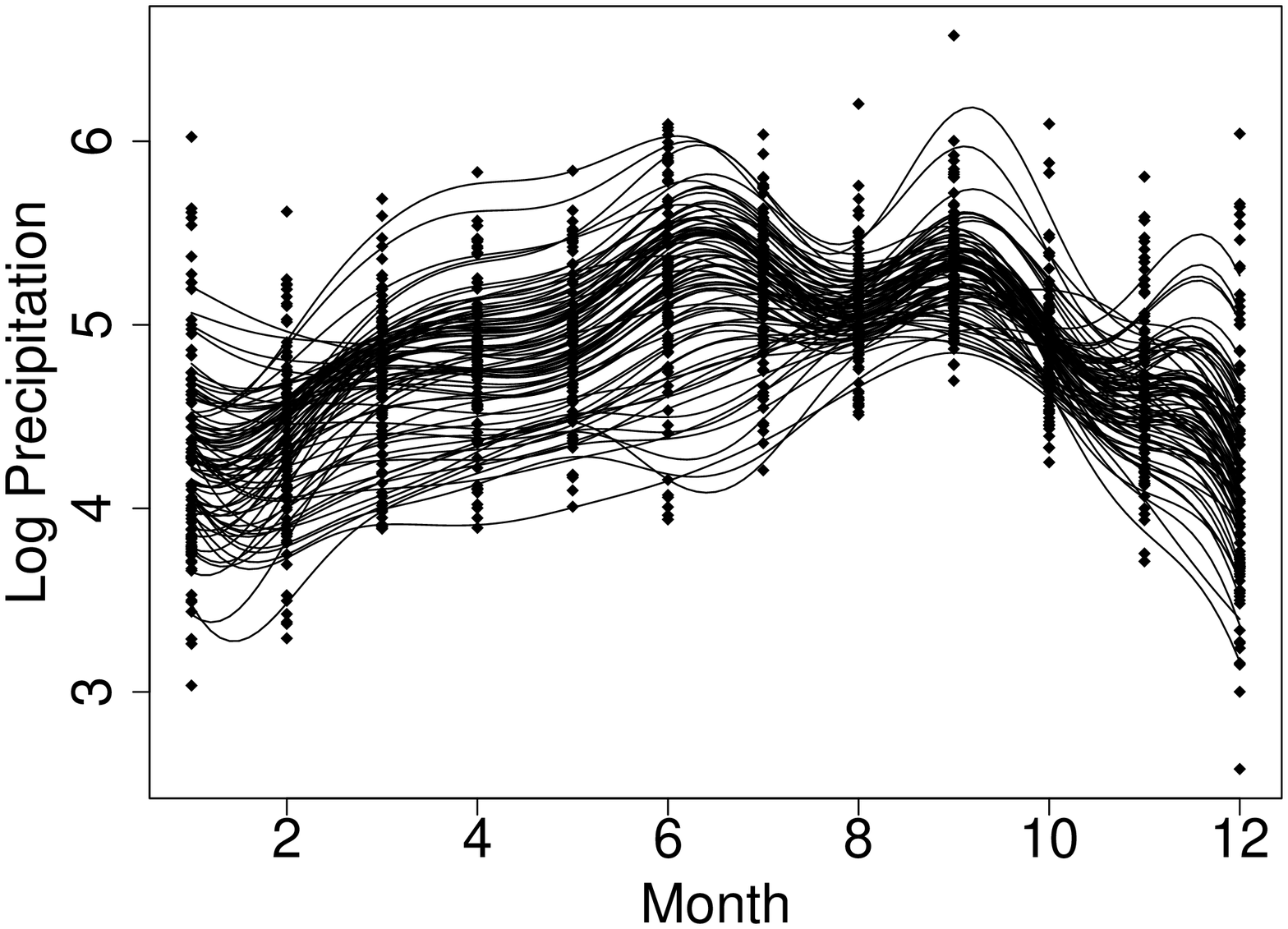}
			\includegraphics[width=7cm]{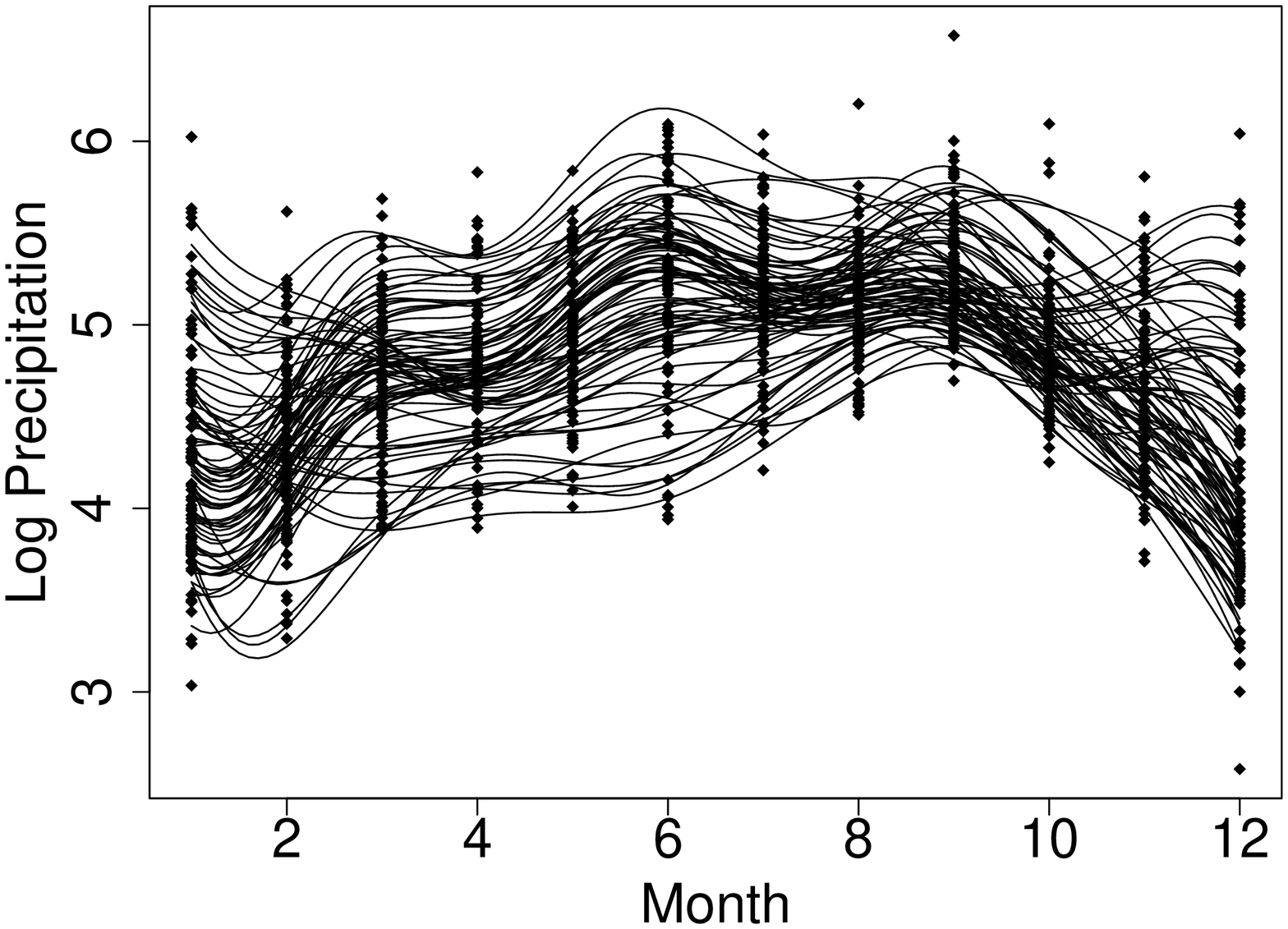}\\		
			\includegraphics[width=7cm]{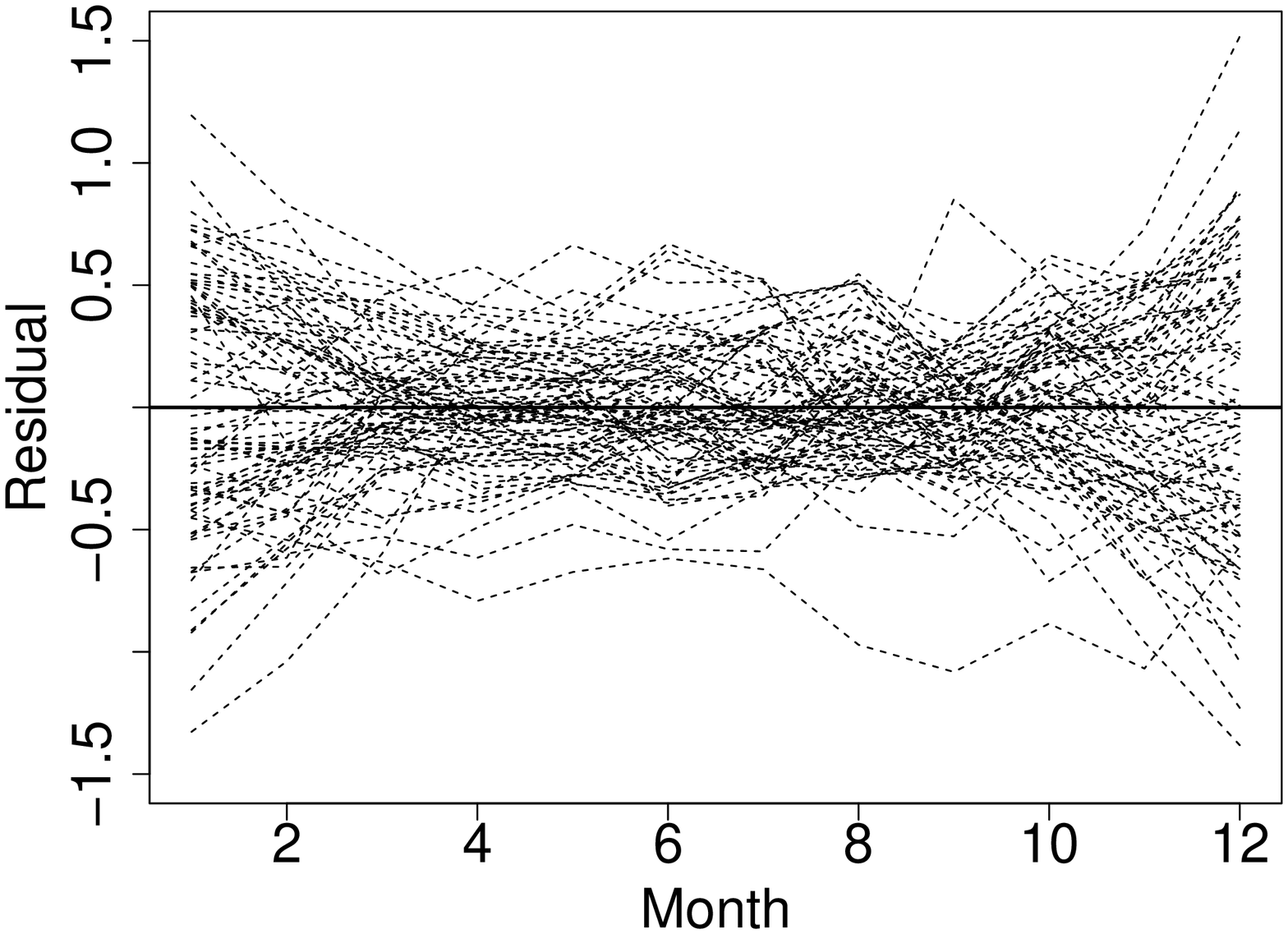}
			\includegraphics[width=7cm]{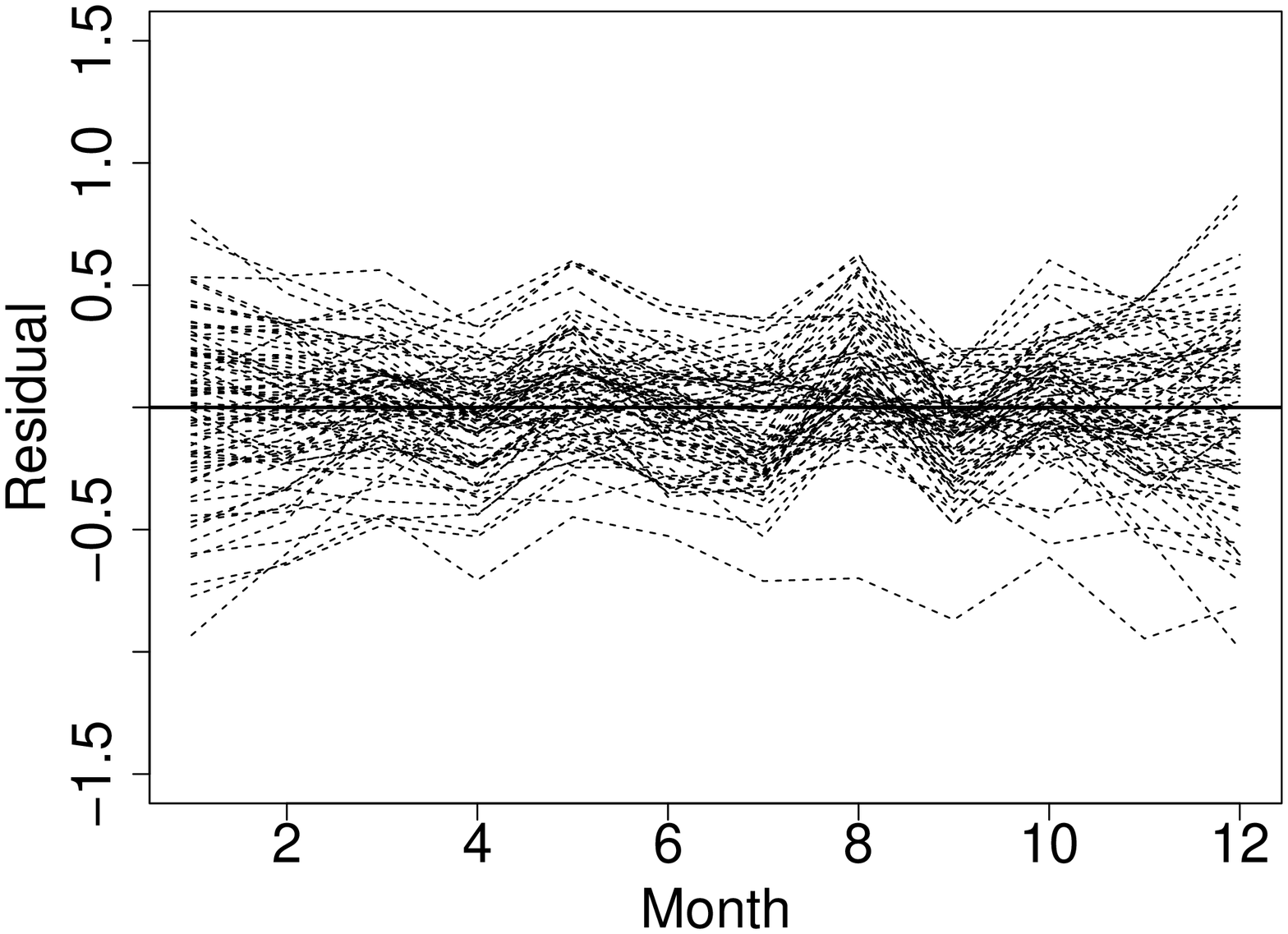}\\		
	\end{center}
	\caption[]{Predicted response curves (Top) and residuals (Bottom) for the model without (Left) and with (Right) the quadratic term.}
	\label{fig:yhat}
\end{figure}

The selected number of basis functions is $M_y=10$ and the selected regularization parameter is $\lambda=1.0\times 10^{-5}$.  
Figure \ref{fig:yhat} compares predicted functions $\hat y_i(t)$ and residuals $y_{ij}-\hat y_i(t_{ij})$ for precipitation obtained from linear and quadratic models.  
Both models broadly capture precipitation trends, but the quadratic model captures individual variation in the response better the linear model.  
The superior performance of the quadratic model is also supposed by the fact that it is associated with smaller residuals compared to the linear model.  
Further, the quadratic model captures individual variations in winter: it can explain the precipitation in cities with heavy snow fall well.  

\begin{figure}[t]
	\begin{center}
			\includegraphics[width=7cm]{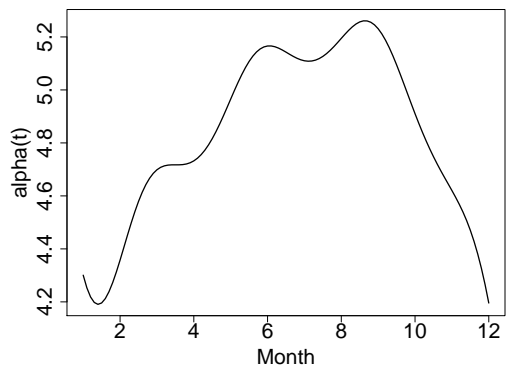} \hspace{1cm}
			\includegraphics[width=7cm]{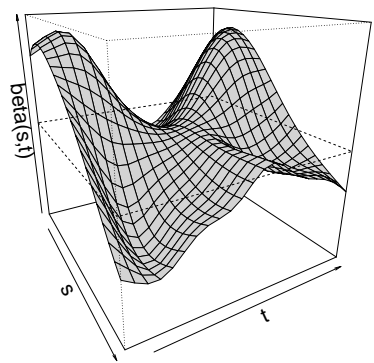}		
	\end{center}
	\caption[]{Left: estimated intercept function $\alpha(t)$.
		Right: estimated coefficient surface for the linear term $\beta(s,t)$. 
		The middle of the vertical axis is zero.}
	\label{fig:coef1}
\end{figure}

\begin{figure}[t]
	\begin{center}
			\includegraphics[width=3.5cm]{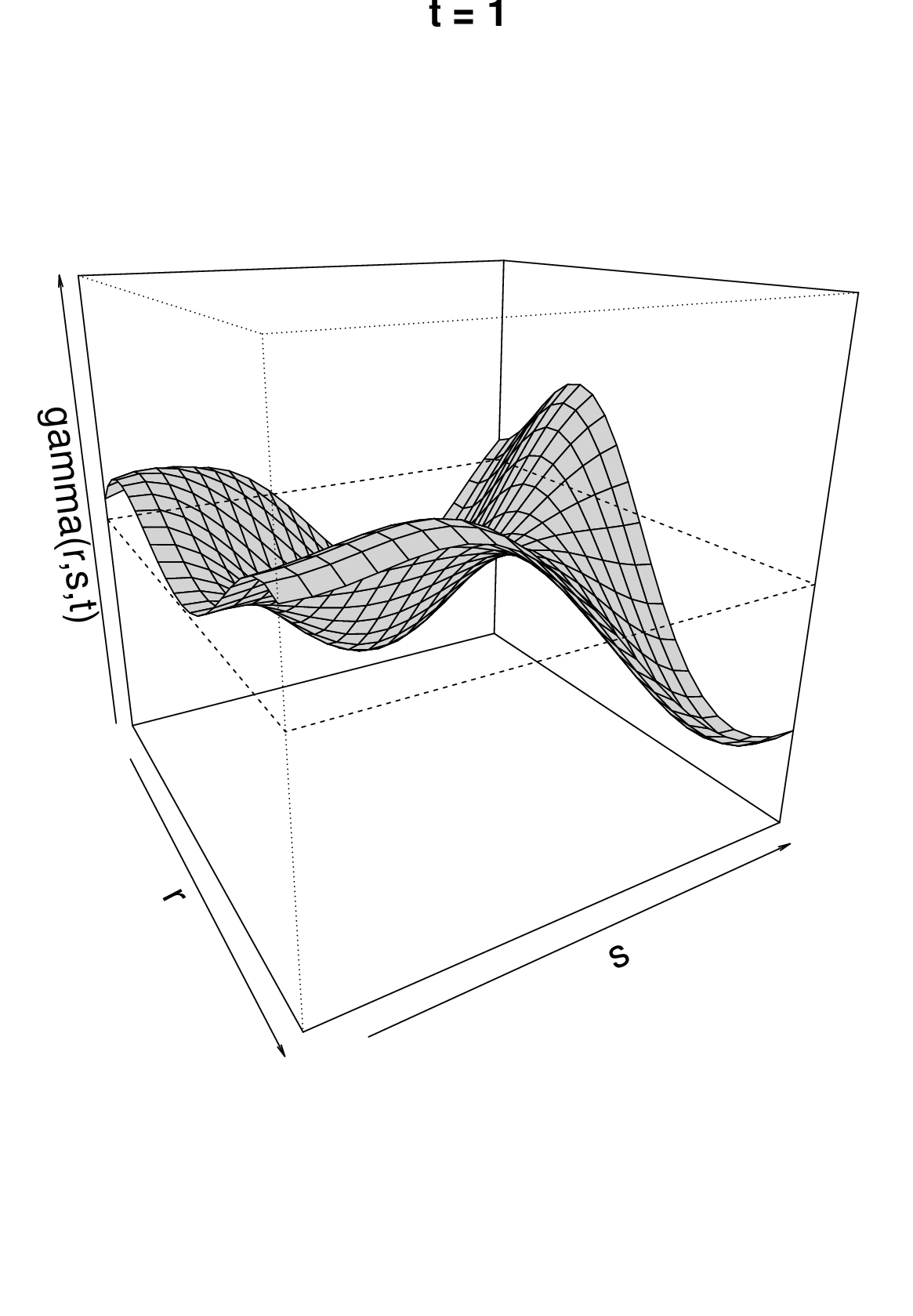}
			\includegraphics[width=3.5cm]{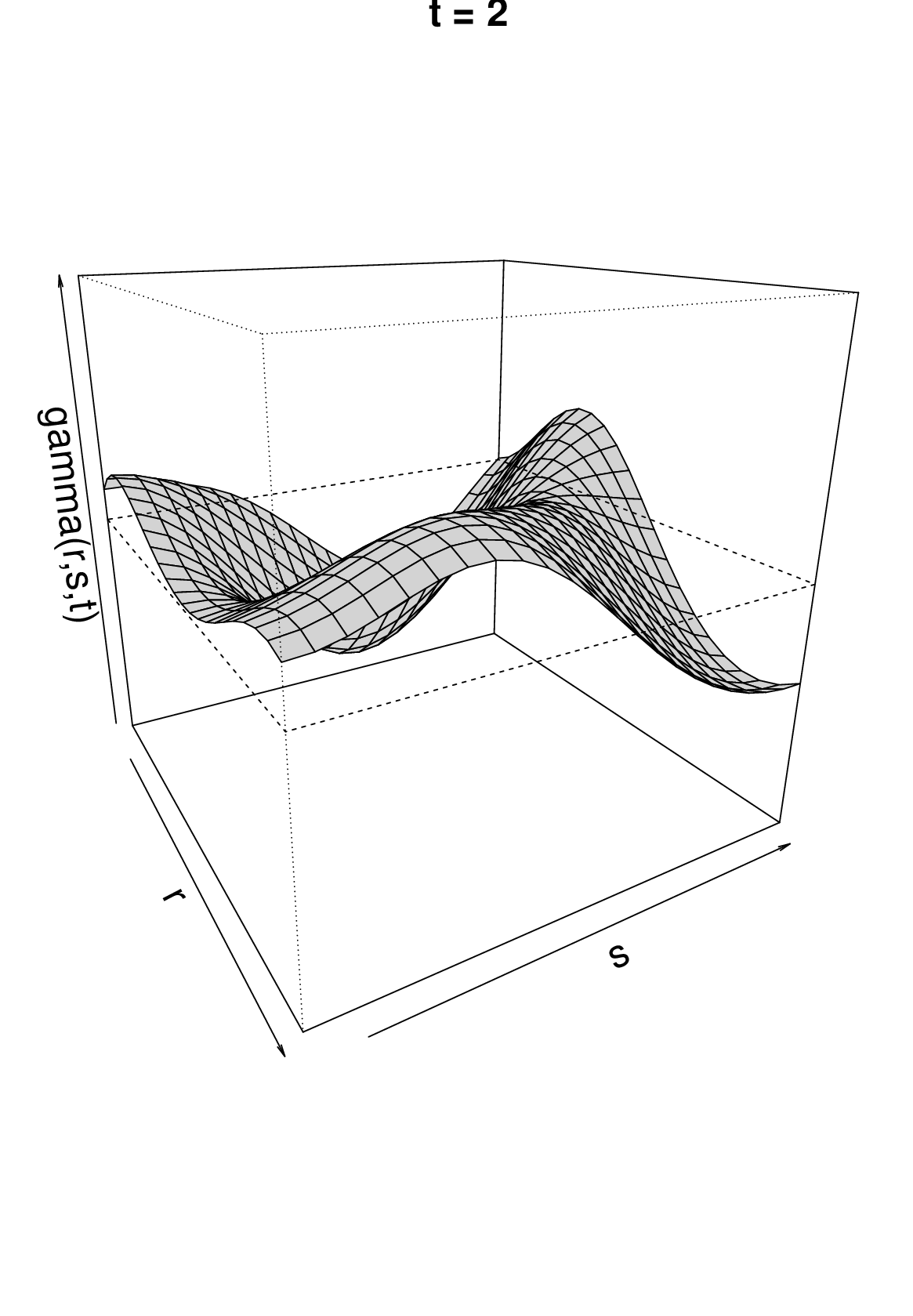}
			\includegraphics[width=3.5cm]{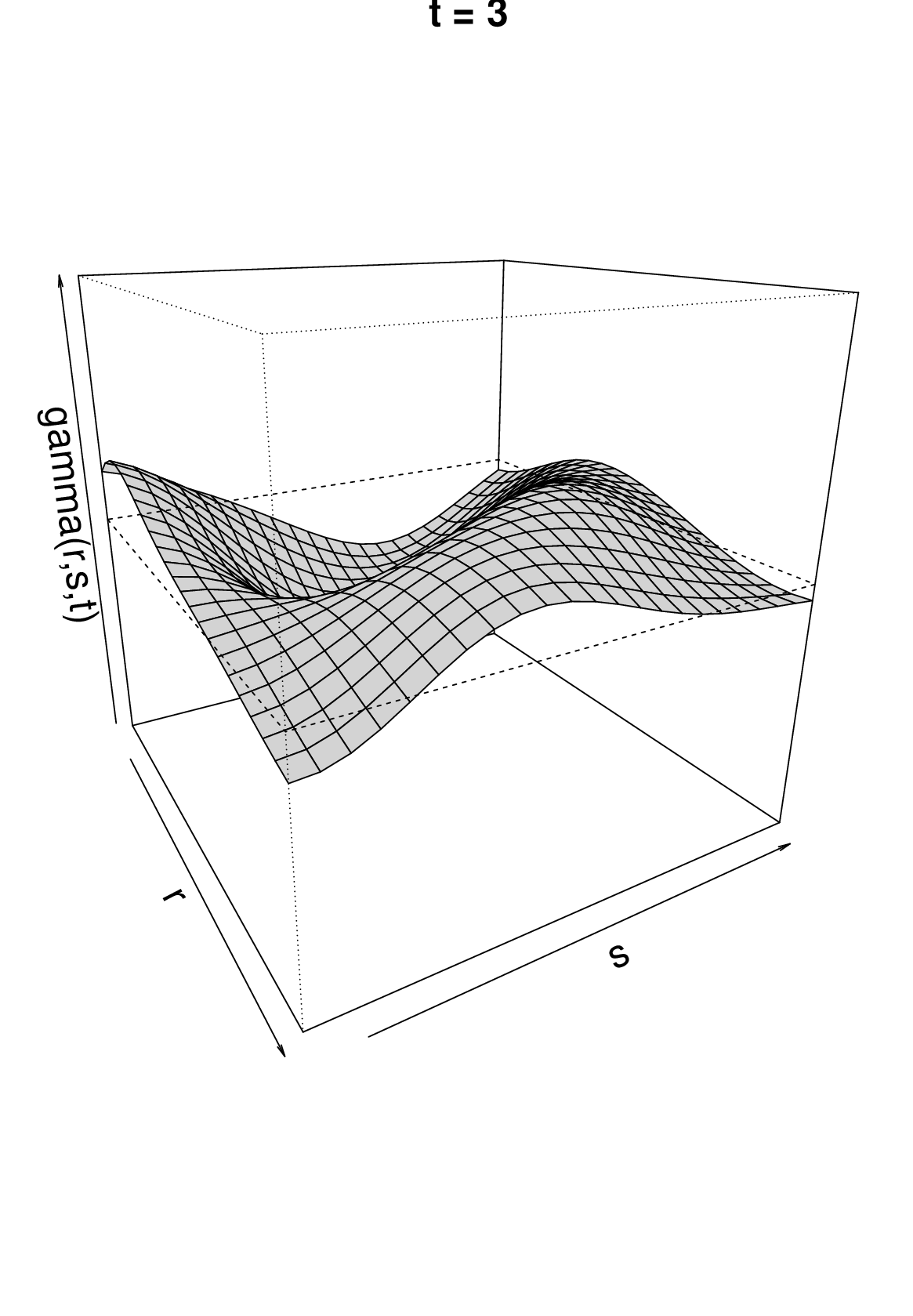}
			\includegraphics[width=3.5cm]{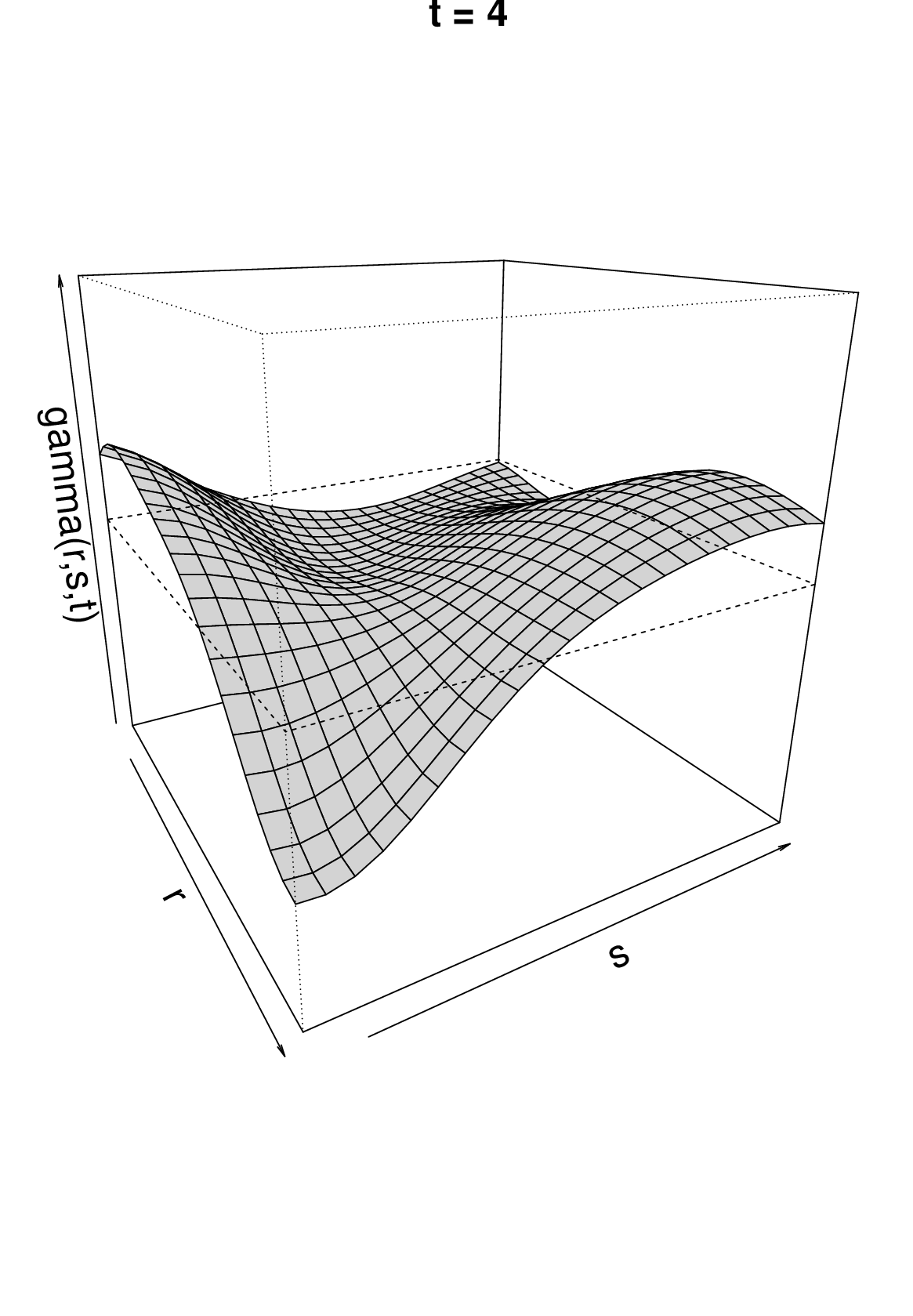}\\
			\includegraphics[width=3.5cm]{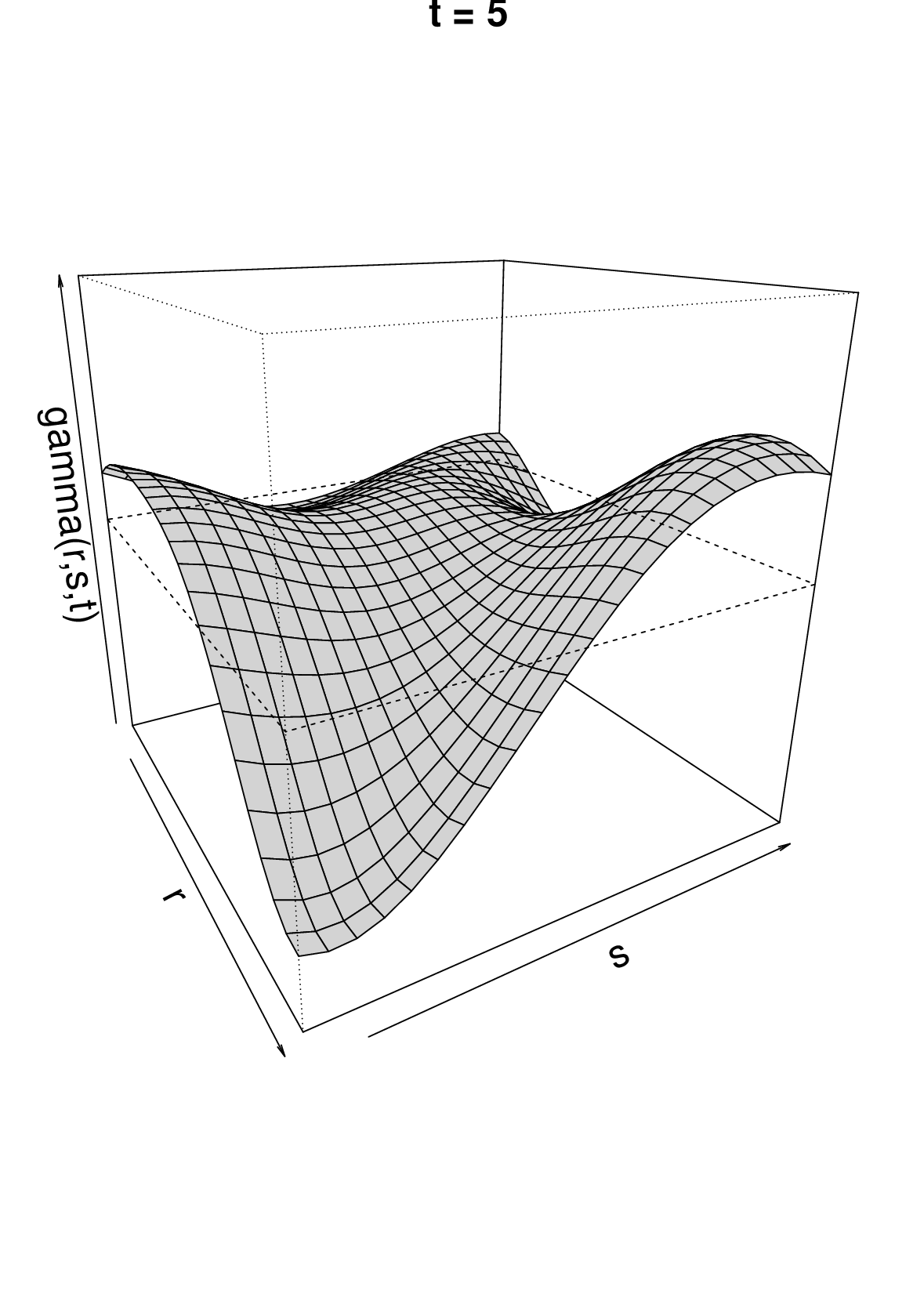}
			\includegraphics[width=3.5cm]{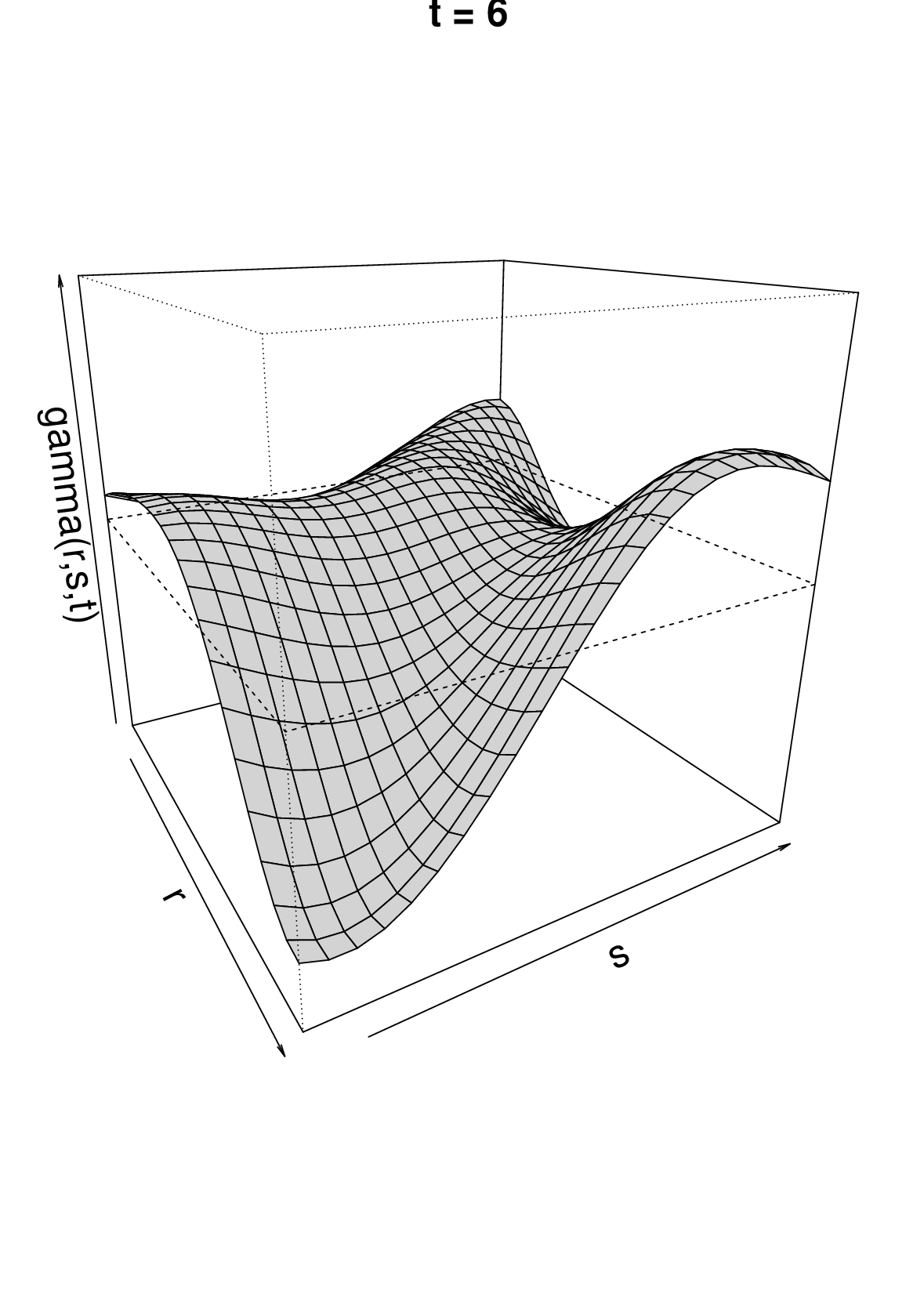}
			\includegraphics[width=3.5cm]{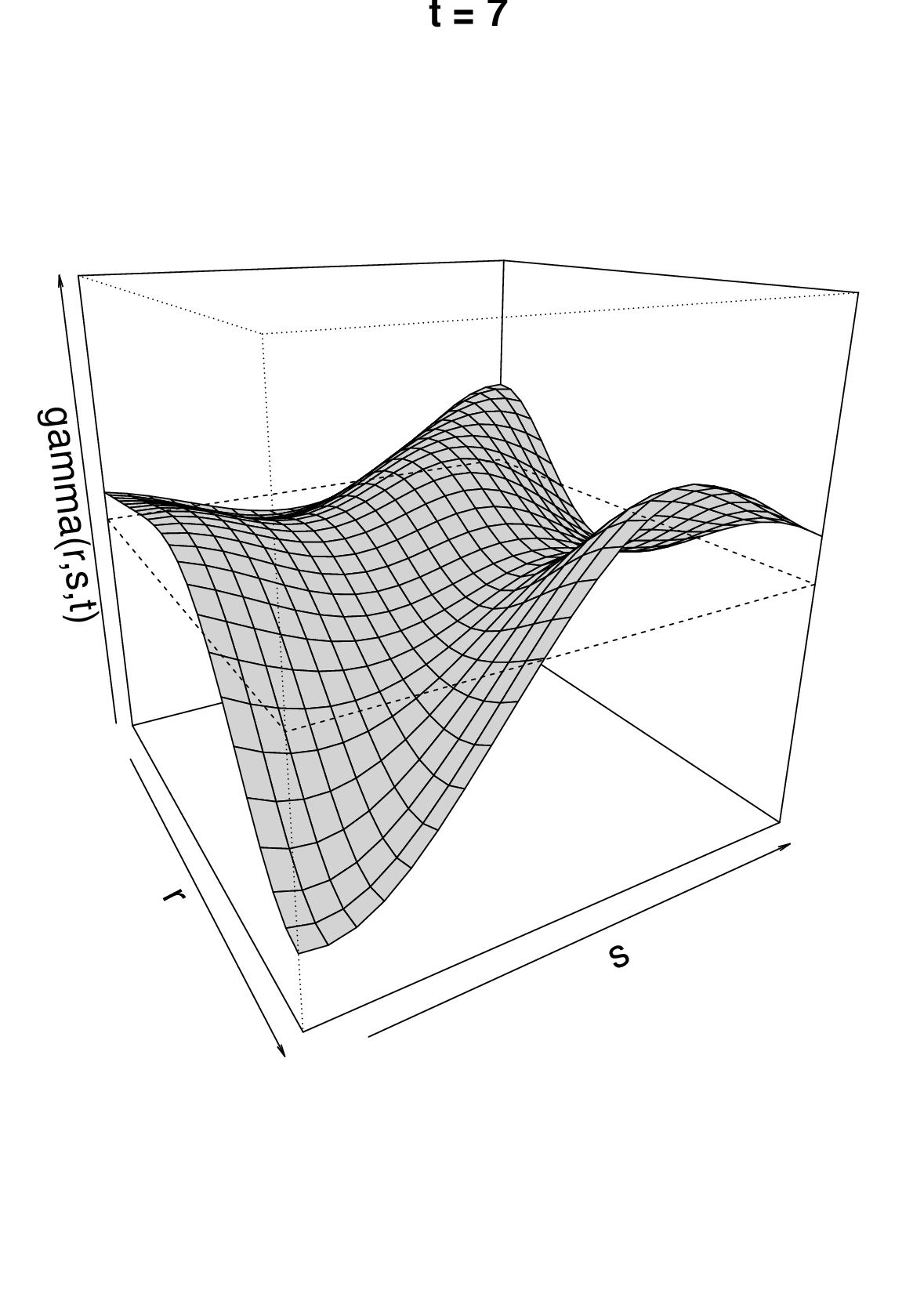}
			\includegraphics[width=3.5cm]{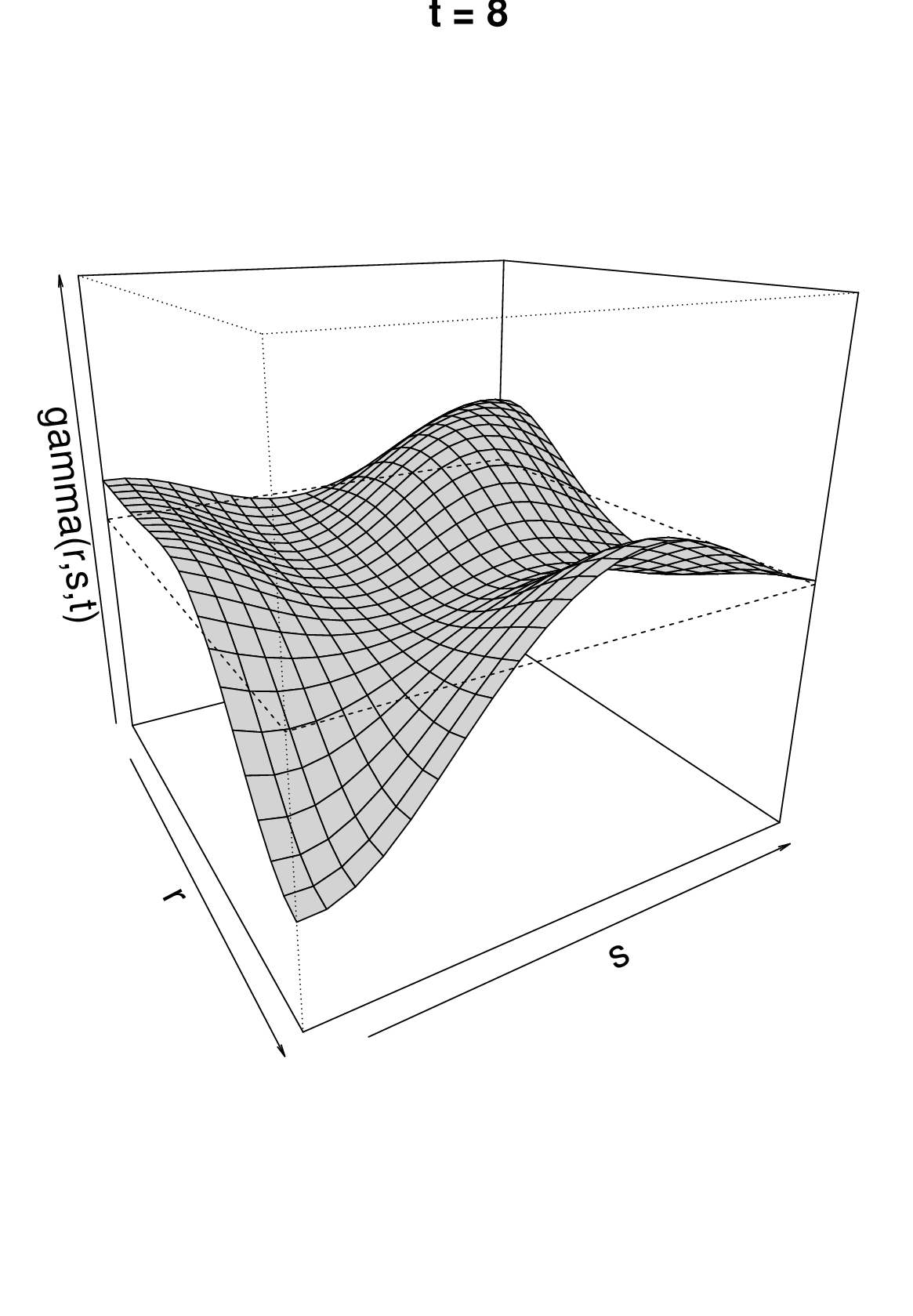}\\
			\includegraphics[width=3.5cm]{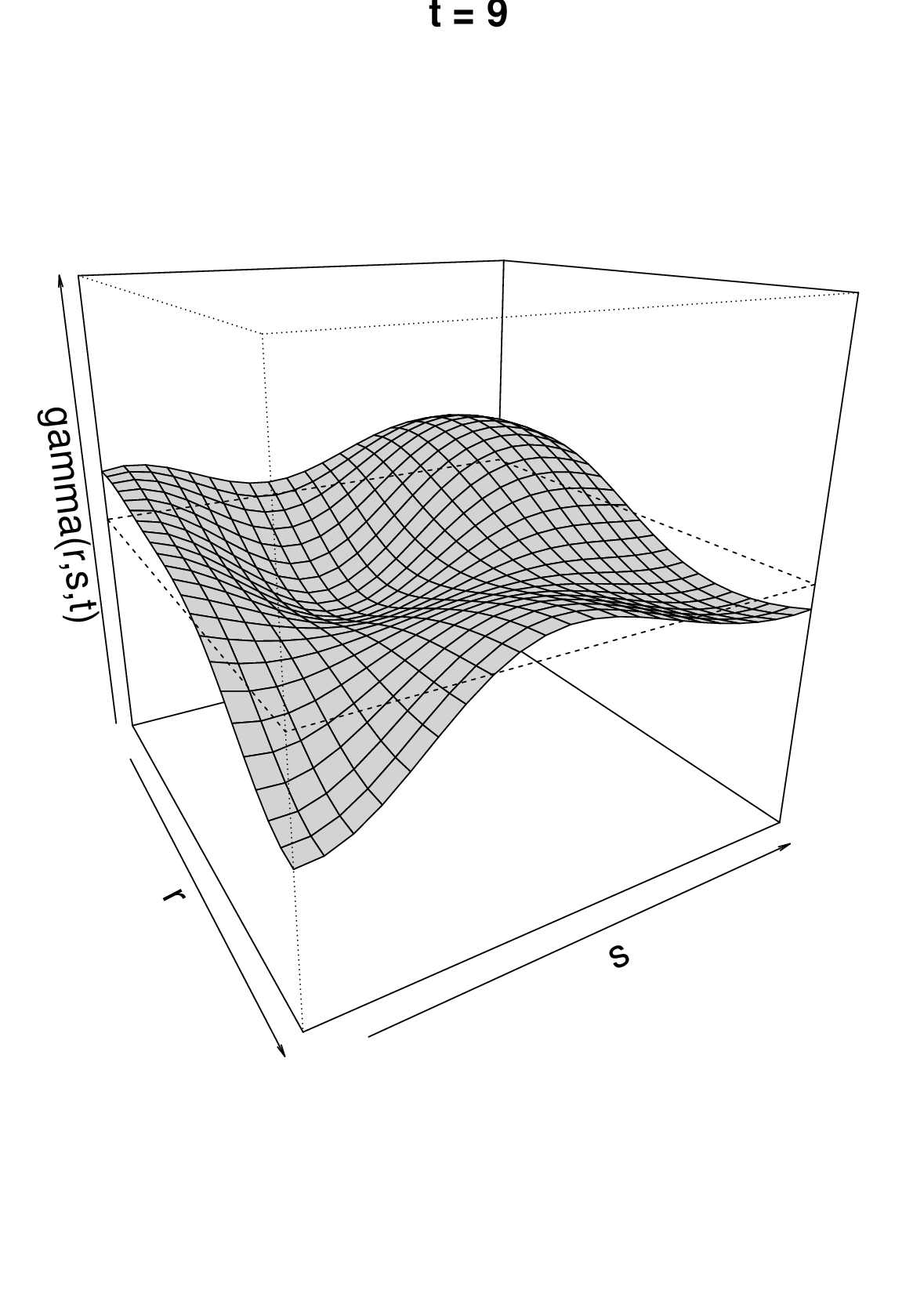}
			\includegraphics[width=3.5cm]{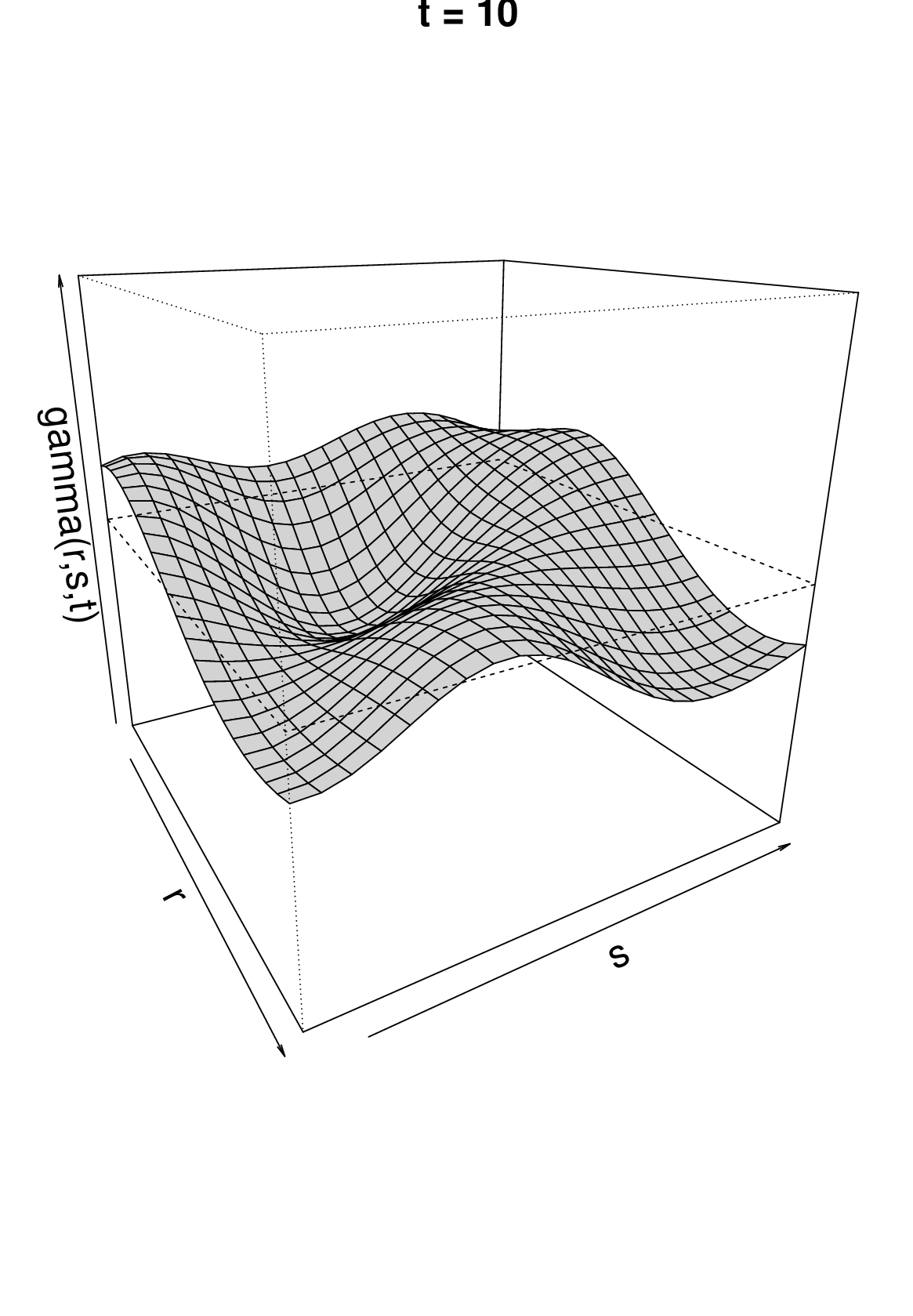}
			\includegraphics[width=3.5cm]{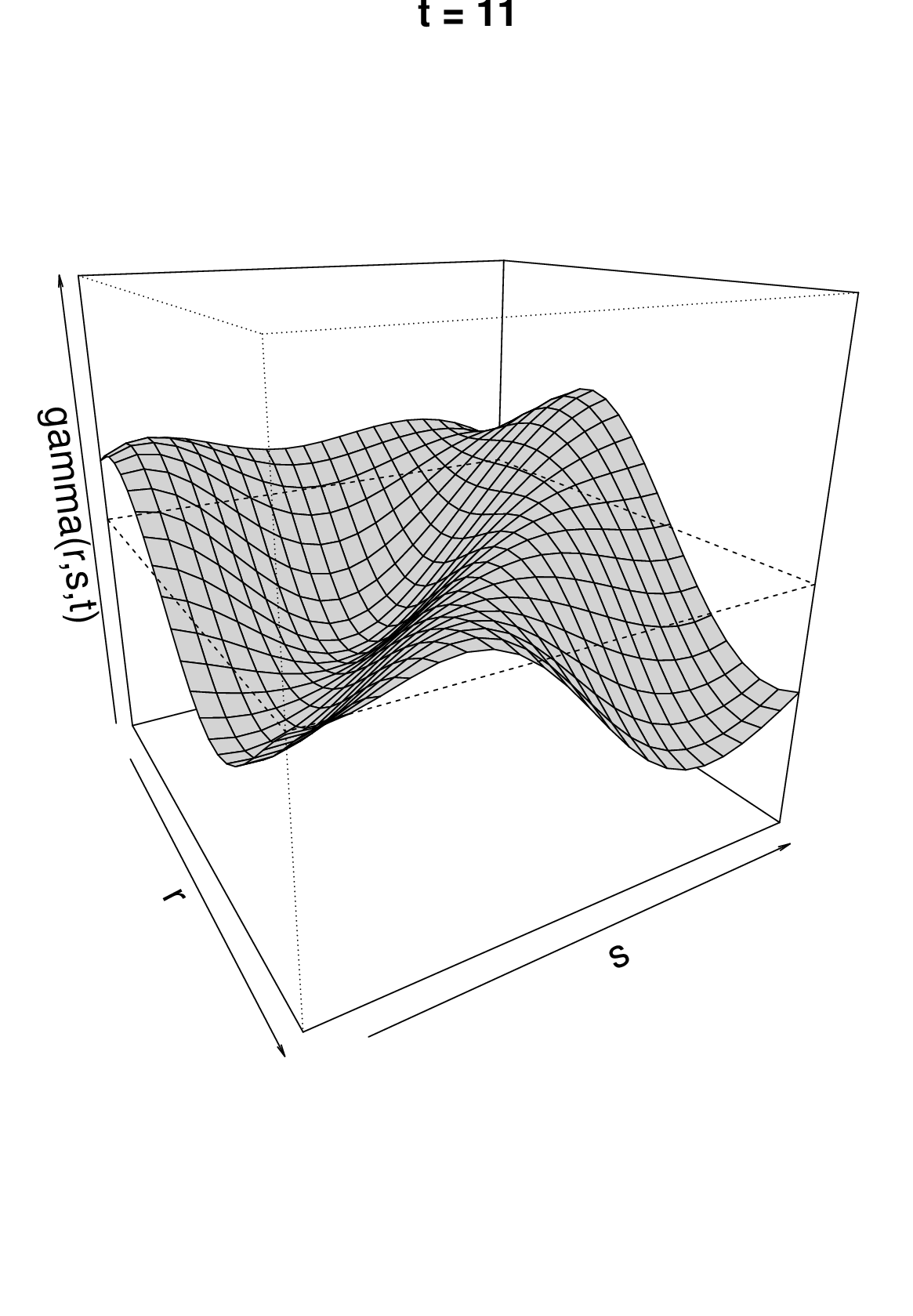}			
			\includegraphics[width=3.5cm]{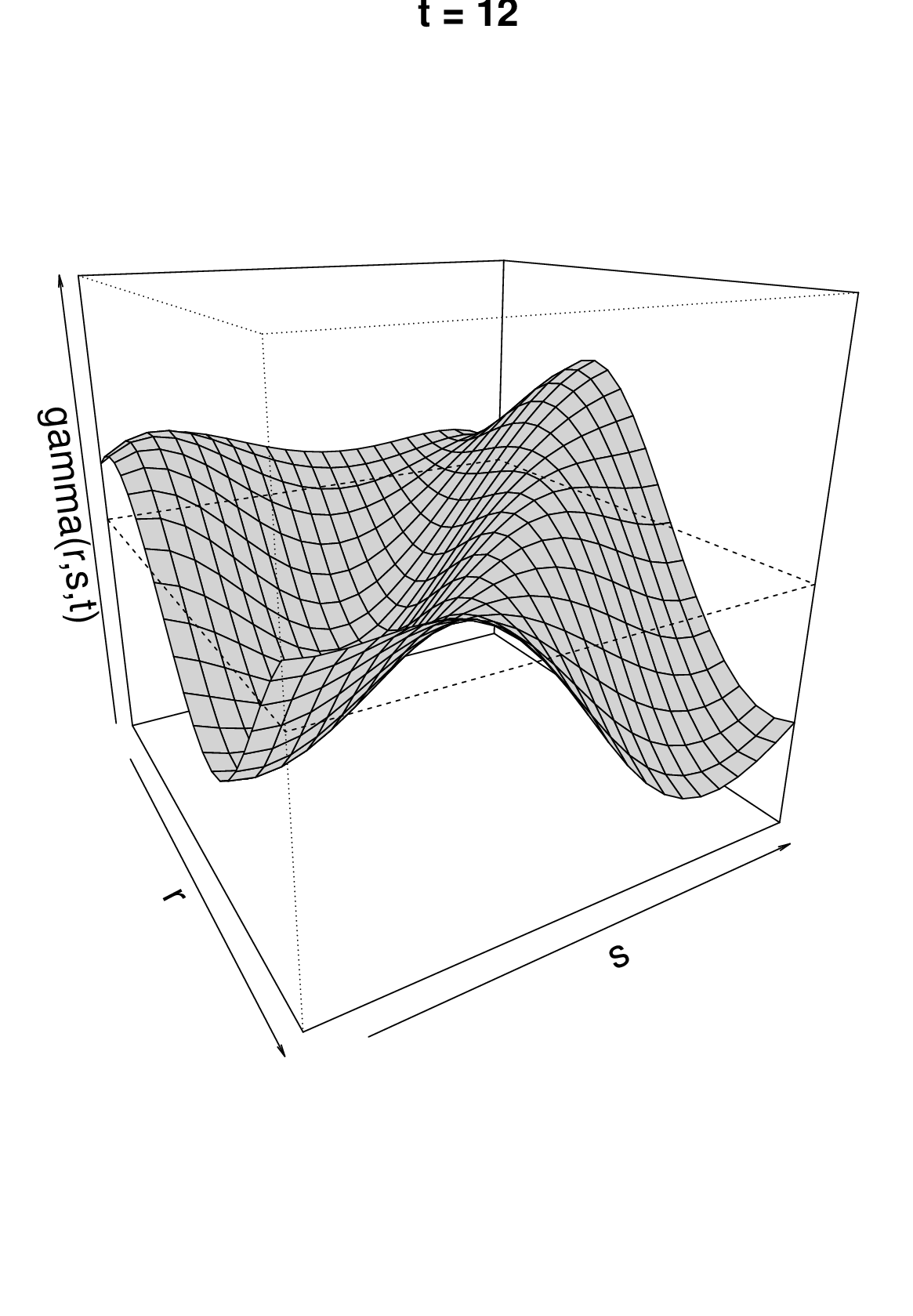}
	\end{center}
	\caption[]{Estimated coefficient hypersurface for the quadratic term $\gamma(r,s,t)$ with $t=1,\ldots,12$ (From top left to bottom right).  The middle of the vertical axis is zero in all graphs.}
	\label{fig:coef2}
\end{figure}

Next, we focus on the estimated functional quadratic model.   
Figure \ref{fig:coef1} shows the estimated baseline and coefficient functions for the linear and quadratic terms.   
The baseline function $\hat\alpha(t)$ broadly captures the overall trend of precipitation functions.
The coefficient surface for the linear term indicates the contribution of temperature $X(s)$ to precipitation $Y(t)$ at arbitrary times $s$ and $t$.  
The result shows that higher temperatures around February contribute to the higher precipitation between fall and spring, whilst higher temperatures at the end of the year contribute to lower precipitation between fall and spring.  
Figure \ref{fig:coef2} shows the estimated hypersurface $\hat\gamma(r,s,t)$ visualized at discrete time points of the precipitation functions.  
This function indicates the interaction of the predictor between at $r$ and $s$ at arbitrary time points $t$.  
For example, we can find that the estimated hypersurface $\gamma(r,s,t)$ has negative weights at the end of $r$ and $s$ for $t=11, 12$, and 1.    
This suggests that lower temperatures around November and December lead to higher precipitation in winter.  
On the other hand, it gives negative weights at the beginning of $s$ and the end of $t$, which indicates that lower temperatures around winter lead to high precipitation between spring and summer.  
\section{Concluding remarks}
Functional regression models elucidate the complex relationship between repeatedly measured variables.  
In this paper, we constructed quadratic regression models for functional data where both the predictor and response are given as functions.  
In addition to representing more flexible relationships between variables than linear models, quadratic variants offer another insight by considering predictor interactions between two time points.  
We assumed Gaussian process for the error function with model parameters estimated by the penalized maximum likelihood method.  
Tuning parameters included in the estimation procedure are selected by model selection criteria.  
The efficacy of our method is evaluated through simulation studies and an empirical example using real data.  

We treated the data as functions of time and assumed annual periodicities.  
In doing so, this generally gives a us the paradoxical insight that future predictor information affects previous responses.  
For this problem, \cite{MaRa2003} proposed a historical functional linear model that takes predictor information only at past response times.  
It would be fruitful for future research to apply a historical functional linear model to our context and analyze data without periodicity.  
For functional linear models with functional response, several estimation methods are proposed beyond what was used in this paper. 
It is thus important to compare the accuracy and computational speed of our method to other methods.  

\section*{Appendix}
We provide details of the derivatives used in our method.  

The first derivatives of $\ell_\lambda(\Theta, \bm{\nu})$ with respect to $\Theta$ and $\nu_j$ are respectively given by 
\begin{align*}
\frac{\partial \ell_\lambda(\Theta, \bm{\nu})}{\partial\Theta} 
&= X^T\Sigma^{-1}(\bm y-X{\rm vec}\Theta) - n\lambda\Omega {\rm vec}\Theta, \\
\frac{\partial \ell_\lambda(\Theta, \bm{\nu})}{\partial \nu_j} 
&= 
\frac{1}{2}{\rm tr}\left\{\left(\bm\alpha\bm\alpha^T - \Sigma^{-1}\right)\frac{\partial \Sigma}{\partial \nu_j}\right\}, \\
\bm\alpha &= \Sigma^{-1}(\bm y - X{\rm vec}\Theta),~~ 
S^{(jj')} = \frac{\partial \Sigma}{\partial \nu_j}\Sigma^{-1}
\frac{\partial \Sigma}{\partial \nu_{j'}}.  
\end{align*} 
The first and second derivatives of the penalized log-likelihood function (\ref{eq:penlike}), used in the Newton-Raphson update (\ref{eq:NR}) and the model selection criteria GIC and GBIC (\ref{eq:IC2}), are given as follows:  
\begin{align*}
\displaystyle{\frac{\partial^2 \ell_\lambda(\Theta, \bm{\nu})}{\partial({\rm vec}\Theta)\partial({\rm vec}\Theta)^T}} 
&= 
-X^T\Sigma^{-1}X - n\lambda\Omega,
\\ 
\displaystyle{\frac{\partial^2 \ell_\lambda(\Theta, \bm{\nu})}{\partial({\rm vec}\Theta)\partial\nu_j}} 
&=
-X^T\Sigma^{-1}\frac{\partial\Sigma}{\partial\nu_j}\Sigma^{-1} (\bm y - X{\rm vec}\Theta),
\\ 
\displaystyle{\frac{\partial^2 \ell_\lambda(\Theta, \bm{\nu})}{\partial\nu_j\partial({\rm vec}\Theta)^T}} 
&= 
\displaystyle{\left(\frac{\partial^2 \ell_\lambda(\Theta, \bm{\nu})}{\partial({\rm vec}\Theta)\partial\nu_j}\right)^T},  \\
\displaystyle{\frac{\partial^2 \ell_\lambda(\Theta, \bm{\nu})}{\partial\nu_j\partial\nu_{j'}}}  
&= 
\frac{1}{2}{\rm tr}\left\{(\bm{\alpha}\bm{\alpha}^T-\Sigma^{-1})
\left(\frac{\partial^2\Sigma}{\partial\nu_j\partial\nu_{j'}} -
S^{(jj')}\right) - \bm{\alpha}\bm{\alpha}^TS^{(jj')}\right\},
\\
%
\displaystyle{\frac{\partial \ell_\lambda(\Theta, \bm{\nu})}{\partial({\rm vec}\Theta)} \frac{\partial \ell(\Theta, \bm{\nu})}{\partial({\rm vec}\Theta)^T}} 
&= 
X^T\bm{\alpha}\bm{\alpha}^TX - n\Omega{\rm vec}\Theta\bm{\alpha}^TX,
\\
\displaystyle{\frac{\partial \ell_\lambda(\Theta, \bm{\nu})}{\partial({\rm vec}\Theta)}\frac{\partial \ell(\Theta, \bm{\nu})}{\partial\bm{\nu}^T}} 
&= 
\frac{1}{2}{\rm tr}\left\{
(\bm{\alpha}\bm{\alpha}^T - \Sigma^{-1})\frac{\partial\Sigma}{\partial\nu_j}
\right\}(X^T\bm{\alpha} - n\lambda\Omega{\rm vec}\Theta),
\\ 
\displaystyle{\frac{\partial \ell_\lambda(\Theta, \bm{\nu})}{\partial\bm{\nu}}\frac{\partial \ell(\Theta, \bm{\nu})}{\partial({\rm vec}\Theta)^T}} 
&= 
\frac{1}{2}{\rm tr}\left\{
(\bm{\alpha}\bm{\alpha}^T - \Sigma^{-1})\frac{\partial\Sigma}{\partial\nu_j}
\right\}X^T\bm{\alpha},
\\
\displaystyle{\frac{\partial \ell_\lambda(\Theta, \bm{\nu})}{\partial\bm{\nu}}\frac{\partial \ell(\Theta, \bm{\nu})}{\partial\bm{\nu}^T}}  
&= 
\frac{1}{4}\left[{\rm tr}\left\{
(\bm{\alpha}\bm{\alpha}^T - \Sigma^{-1})\frac{\partial\Sigma}{\partial\nu_j}
\right\}\right]^2.
\end{align*}
If we assume a Gaussian process with a Gaussian covariance function as in (\ref{eq:GP}), the first and second derivatives of $\Sigma_i$ with respect to $\nu_j$ $(j=1,2,3)$ are given by
\begin{align*}
\frac{\partial \Sigma_i}{\partial \nu_1} 
&= 
\left(\exp\left\{-\frac{\nu_2}{2}(t_{ij}-t_{ij'})^2\right\}\right)_{jj'}, \\
\frac{\partial \Sigma_i}{\partial \nu_2} 
&= 
\left(-\frac{\nu_1}{2}(t_{ij}-t_{ij'})^2\exp\left\{-\frac{\nu_2}{2}(t_{ij}-t_{ij'})^2\right\}\right)_{jj'}, \\
\frac{\partial \Sigma_i}{\partial \nu_3} &= I_{n_i},\\
\frac{\partial^2 \Sigma_i}{\partial \nu_1^2} 
&= 
\frac{\partial^2 \Sigma_i}{\partial \nu_1\partial \nu_3}
=  
\frac{\partial^2 \Sigma_i}{\partial \nu_2\partial \nu_3} 
=  
\frac{\partial^2 \Sigma_i}{\partial \nu_3^2} 
= O, \\
\frac{\partial^2 \Sigma_i}{\partial \nu_1\partial \nu_2} 
&= \left(-\frac{1}{2}(t_{ij}-t_{ij'})^2\exp\left\{-\frac{\nu_2}{2}(t_{ij}-t_{ij'})^2\right\}\right)_{jj'}, \\
\frac{\partial^2 \Sigma_i}{\partial \nu_2^2} 
&= \left(-\frac{\nu_1}{4}(t_{ij}-t_{ij'})^2\exp\left\{-\frac{\nu_2}{2}(t_{ij}-t_{ij'})^2\right\}\right)_{jj'}.
\end{align*}

\end{document}